\documentclass[10pt,a4paper,onecolumn,oneside]{article}
\usepackage[latin1]{inputenc}

\usepackage[
a4paper,
asymmetric,                               
textwidth=16cm,
outer=2.5cm,
textheight=45\baselineskip,
headheight=\baselineskip,
includehead=true,
heightrounded,
]{geometry}

\usepackage{ifthen}

\usepackage{fancyhdr}
\usepackage{soul}
\usepackage{graphicx} 
\usepackage{natbib} 
\usepackage{amsmath} 
\usepackage{amssymb} 
\usepackage{amsbsy} 
\usepackage{amsgen} 
\usepackage{amsfonts} 
\usepackage{pgfbaseimage}
\usepackage{xspace}
\usepackage{psfrag}
\usepackage{import}
\usepackage{booktabs}
\usepackage[pscoord]{eso-pic}
\usepackage{tikz} 
\usetikzlibrary{snakes} 
\usetikzlibrary{patterns}										 
\usepackage{siunitx}
\usepackage{units}
\usepackage{float}

\DeclareMathOperator\erf{erf}
\DeclareMathOperator\erfc{erfc}

\newcommand\mpers{\nobreak\mbox{$\:$m$\:$s$^{-1}$}}

\newcommand\Pn{P\'eclet number\xspace}
\newcommand\Pns{P\'eclet numbers\xspace}

\newcommand\Rey{\mbox{\textit{Re}}}  
\newcommand\Sc{\mbox{\textit{Sc}}} 
\newcommand\Sh{\mbox{\textit{Sh}}}
\newcommand\Pe{\mbox{\textit{Pe}}}
\newcommand\Nu{\mbox{\textit{Nu}}}
\newcommand\ie{i.e.\xspace}
\newcommand\eg{e.g.\xspace}
\newcommand\p{\ensuremath{\partial}}
\newcommand\dd{\ensuremath{\mathrm{d}}}

\providecommand\bnabla{\boldsymbol{\nabla}}
\providecommand\bcdot{\boldsymbol{\cdot}}

\renewcommand{\(}{\left(}
\renewcommand{\)}{\right)}
\renewcommand{\[}{\left[}
\renewcommand{\]}{\right]}

\DeclareMathAlphabet{\mathpzc}{OT1}{pzc}{m}{it}

\begin{document}

\thispagestyle{plain}


\begin{center}
	\textbf{ \Large  Convective mass transfer from a submerged drop in a thin falling film \\ \vspace{0.7cm}
	\normalsize  \sc Julien R. Landel$^1$, \ A. L. Thomas$^2$, \ H. McEvoy$^3$ \ {\normalfont and} \ Stuart B. Dalziel$^1$} \\ \vspace{0.35cm}
	\small $^1$Department of Applied Mathematics and Theoretical Physics, University of Cambridge,
		Centre for Mathematical Sciences, Wilberforce Road, Cambridge, CB3 0WA, UK \\ \vspace{0.1cm}
	$^2$Department of Physical and Environmental Sciences, National University of Central Buenos Aires, Tandil, Buenos Aires Province, Argentina\\ \vspace{0.1cm}
	$^3$Defence Science and Technology Laboratory, Salisbury, Wiltshire, SP4 0JQ, UK	 \\ \vspace{0.35cm}
	\today \\ \vspace{0.5cm}
	Approved by DTRA for public release, distribution is unlimited. \\ \vspace{0.35cm}
\end{center}

\begin{abstract}
We investigate the fluid mechanics of removing a passive tracer contained in small, thin, viscous drops attached to a flat inclined substrate using thin gravity-driven film flows. We focus on the case where the drop cannot be detached either partially or completely from the surface by the mechanical forces exerted by the cleaning fluid on the drop surface. Instead, a convective mass transfer establishes across the drop--film interface and the dilute passive tracer dispersed in the drop diffuses into the film flow, which then transports them away. The \Pn  for the passive tracer in the film phase is very high, whereas the \Pn in the drop phase varies from  {$\Pe_d \approx 10^{-2}$} to $1$. The characteristic transport time in the drop is much larger than in the film. We model the mass transfer of the passive tracer from the bulk of the drop phase into the film phase using an empirical model based on an analogy with Newton's law of cooling. This simple empirical model is supported by a theoretical model solving the quasi-steady two-dimensional advection--diffusion equation in the film coupled with a time-dependent one-dimensional diffusion equation in the drop. We find excellent agreement between our experimental data and the two models, which predict an exponential decrease in time of the tracer concentration in the drop. The results are  {valid for all  drop and film \mbox{\Pns} studied}. The overall transport characteristic time is related to the drop diffusion time scale, as diffusion within the drop is the limiting process. This result remains valid even for $\Pe_d \approx 1$. Finally, our theoretical model predicts the well-known relationship between the  Sherwood number and the  Reynolds number in the case of a well-mixed drop  {$\Sh \propto \mbox{$Re_L$}^{1/3}=(\gamma L^2/\nu_f)$, based on the drop length $L$, film shear rate $\gamma$ and film kinematic viscosity $\nu_f$. We show that this relationship is mathematically equivalent to a more physically intuitive relationship $\Sh \propto Re_{\delta}$, based on the diffusive boundary layer thickness $\delta$.} The model also predicts a correction in the case of a non-uniform drop concentration. The correction depends on $Re_{\delta}$, the film Schmidt number, the drop aspect ratio and the diffusivity ratio between the two phases. This prediction is in remarkable agreement with experimental data at low drop \Pn. It continues to agree as  $\Pe_d$ approaches $1$, although the influence of the  Reynolds number increases such that $\Sh \propto Re_{\delta}$.
\end{abstract}

\section{Introduction}

In the present study, we are interested in removing material contained in small viscous droplets lying on an inclined substrate using thin falling films. We focus on the case where the droplets cannot be detached from the substrate. The shear and pressure forces imposed by the film flow on the droplet cannot overcome the adhesive forces between the droplet and the substrate, and the cohesive forces within the droplet.  {Instead, the cleaning or removal of the passive tracer contained in the droplet occurs through a continuous mass transfer  from the droplet into the submerging film. The tracer is then transported away by the film flow.}

\begin{figure}[H]
	\centering
	\scriptsize
	\def\svgwidth{0.5\textwidth}
	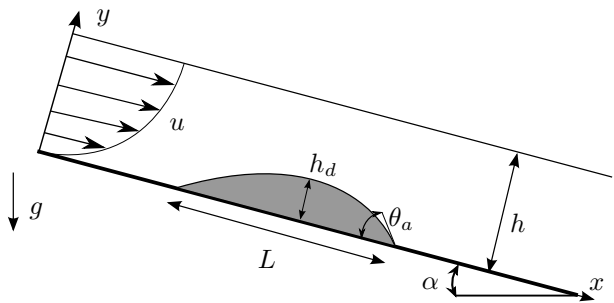
	\caption{Schematic diagram of the cleaning problem or convective mass  {transfer  at large film \mbox{\Pns}. A thin liquid film driven by gravity flows over a small  viscous droplet, modelled as a thin spherical cap,} containing some dilute passive tracer (shaded).}
	\label{fig:droplet-film}
\end{figure}

Our aim is to understand and model the fundamental physical phenomena which govern the convective mass transfer from the droplet into the film. We analyse, through experimental investigation and theoretical modelling, the mass transfer of a dilute passive tracer dispersed in a single droplet, which is submerged in a liquid film (see figure~\ref{fig:droplet-film}). The typical size of the droplet is approximately $1\;$mm in diameter at its base and a few hundred microns in thickness. The film falling over the droplet is approximately $1\;$mm in thickness, which is slightly larger than the drop  {thickness.}

Cleaning through diffusion of small viscous droplets spread over a substrate using thin falling films is a common problem in many industrial applications. This is particularly important in the food industry to ensure high hygiene standards \cite[see \eg][]{wilson05}. As \cite{landel15} pointed out, this problem   {also exists} in our daily life when using dishwashers. In a dishwasher, most surfaces of the dishes cannot be cleaned through the mechanical forces of a strong liquid jet as they are often not directly accessible to the jet. Instead, a draining   {film  covers} all surfaces and removes grease or other food deposits through a slow diffusion process.   {This cleaning method tends to reduce water and energy consumption, and the volume of waste, due to lower wetting rates \mbox{\cite[][]{fuchs13}}.} The present study is also applicable to the decontamination of  {hazardous substances mixed in viscous liquid droplets \mbox{\cite[][]{fatah07,danberg08}}. A liquid decontaminant can be delivered onto contaminated surfaces flowing over the hazardous droplets as a thin falling film.}  The cleaning of cars that have been fouled with drops of tree resin can also require the dissolution of the resin in a cleaning film flow.
Other applications can concern the evaporation of drops in a shear flow: for spray coating, rain drops on a wind screen, or spraying of chemicals on crops \cite[][]{baines:james94}.

The problem of convective mass transfer from a small droplet into a surrounding flowing fluid has been approached in different ways in the literature. In the external phase (\ie the film phase in our problem), the problem can be seen as an advection--diffusion problem with imposed Dirichlet or Neumann  boundary conditions at the interface between the two phases.  {Also, the mass transfer considered in this  study is equivalent to the problem of heat transport, as the low concentration of passive tracer involved is in the dilute regime  \mbox{\cite[\eg][]{kays05}}.} 

The mathematical derivation of the two-dimensional advection--diffusion equation in the case of a developed shear flow over a rigid substrate with Dirichlet boundary  {conditions is} presented in \cite{bejan13}. Asymptotic solutions of the Nusselt number $\Nu$ are given as a function of the length of the heated boundary, the Reynolds number $\Rey$ and the Prandtl number $Pr$. The Nusselt number, $\Nu=\lambda \ell/D$ (with $\lambda$ the convective heat transfer coefficient, $\ell$ a characteristic length scale and $D$ the diffusion coefficient), characterizes the importance of convective processes for the heat transfer in comparison with diffusive processes. The equivalent of the Nusselt number for mass transfers, and one of the key points studied in this paper, is the Sherwood number, $\Sh=\kappa \ell/D$, where $\kappa$ is the convective mass transfer coefficient. Typically, the Sherwood number, averaged over the length of the plate, varies with Reynolds number and Schmidt number as $\Sh \propto \Rey^{1/3} \Sc^{1/3}$, at large Schmidt numbers. This result is based on the pioneering work of \cite{leveque28}, who assumed that the velocity profile in the diffusive boundary layer was linear (\ie a uniform shear flow). This assumption holds true at large Schmidt numbers because the momentum boundary layer grows faster than the diffusive boundary layer.

\cite{stone89} studied the heat or mass transfer from a flat patch of arbitrary shape into a two-dimensional developed shear flow. He found good agreement between his numerical simulations and his corrected asymptotic solution for the Nusselt number $\Nu= 2.157 \Pe^{1/3}+3.55\Pe^{-1/6}$, at large  \Pns ($\Pe=\Rey Pr=U\ell/D>10$, with $U$ a characteristic velocity in the flow). The correction term $\Pe^{-1/6}$ accounts for edge effects at the patch boundary, where the heat flux is discontinuous. 

\cite{baines:james94} studied theoretically the specific case of a flat liquid drop of water evaporating into a gas shear flow at large external \Pns. They derived the two-dimensional asymptotic solution of the Sherwood number averaged over the drop surface, $\Sh = 0.105 \Pe^{1/3} $, by solving the advection--diffusion equation using a similarity solution. In their model, they assumed that the droplet is completely flat and does not affect the film flow. The comparison of their theoretical prediction with the experimental data of \cite{coutant:penski82} shows a very poor agreement. \cite{baines:james94} noted a very large scattering in the experimental data, which they attributed to the sensitivity of the measurements of the drop contact angle and the lack of consistency between the different experiments of  \cite{coutant:penski82}.

\cite{danberg08} studied experimentally, numerically and theoretically the same problem as \cite{baines:james94}: the evaporation of a small flat non-deformable liquid drop  into a gas shear flow at large  \Pns. He found a similar theoretical prediction for the Sherwood number using an integral method and assuming a third-order polynomial distribution for the concentration in the diffusive boundary layer. His numerical simulation of the two-dimensional advection--diffusion equation agrees well with his asymptotic solution of the Sherwood number, $\Sh \propto \Pe^{1/3} $. The experimental results of \cite{danberg08} show a better agreement than the results of \cite{baines:james94} with the asymptotic solution of the Sherwood number.  {However, there is still some scattering in the experimental data for the Sherwood number, with $\pm 17$\%  deviation for the $95$\% confidence interval on the average experimental slope. The experimental slope is also reported to be $13$\% larger than the predicted slope.}

It is clear from the works of \cite{coutant:penski82}, \cite{baines:james94} and \cite{danberg08} that the experimental measurements of the mass transfer from a single droplet into a surrounding shear flow are very challenging. In fact, the theoretical prediction $\Sh \propto \Pe^{1/3} $ still has to be confirmed experimentally.
 We also note that in all three studies \cite[][]{coutant:penski82,baines:james94,danberg08}, and contrary to the present study, the drops considered contained only one species. This case tends to simplify the evaluation of the overall mass transfer because it fixes the concentration at the interface: the concentration at the interface is constant in time and exactly equal to the concentration at any point inside the drop. Thus, they did not need to study the mass transport in the drop phase. The overall mass transfer could be determined completely from the study of the advection--diffusion equation in the external phase, following the mathematical derivation summarized in \cite{bejan13}.
 
 In our study, we investigate the more complex case of a drop constituted of several species. The passive tracer, which is transported out of the droplet, has a small concentration and can be considered  dilute in the drop. The volume of the drop can thus be considered  constant in time. Having different species inside the drop requires the study of the transport of the tracer inside the drop and across the interface. The transport in the external phase is coupled with the internal transport through the boundary condition at the interface. Furthermore, in our problem the drop is submerged in a film of finite thickness, instead of an infinite medium with a uniform shear flow. \cite{landel15} studied experimentally the impact of a rigid sessile drop on the velocity field in thin falling films. This work showed a significant decrease of the streamwise velocity at the surface of the film and a complex three-dimensional return flow immediately downstream of the drop. We also consider in the present study the impact of the deformation of the drop shape by the film, although the viscous drop  deforms at a comparatively slow rate under the action of the film shear forces  {\mbox{\cite[see \eg][]{fan11}}}.

Our main goal is to understand and predict mathematically how the concentration of a dilute chemical component or passive tracer in a deformable drop changes in time as a thin film flows over it. In section \ref{sec:mathmodel}, we first review the physical phenomena and associated characteristic time scales governing these transport processes. In particular, we note that transport in the film phase is strongly dominated by advection, since the film \Pn is very high, whereas transport in the drop phase can be either dominated by diffusion at low drop \Pns, or a combination of both advection and diffusion for $\Pe_d\approx 1$. Next, we model the overall mass transfer, which predicts the time variation of the tracer concentration in the droplet, using a simple empirical model based on an analogy with Newton's law of cooling. We then propose a coupled model based on fundamental physical principles to support the Newton-cooling model and derive a theoretical prediction for the Sherwood number. We model transport in the film phase by solving a quasi-steady advection--diffusion equation in the diffusive boundary layer, following previous works \cite[][]{blount10,bejan13}. In the drop phase we use a one-dimensional time-dependent diffusion model for the transport at low drop \Pns. In section~\ref{sec:xpproc}, we describe our experimental procedure. In particular,  {we describe} the dye attenuation technique, which is used to measure the flux of tracer, methylene blue dye, from the droplet into the submerging film. In section~\ref{sec:xpresults}, we compare our experimental results with our theoretical predictions for the temporal evolution of the tracer concentration in the drop and the Sherwood number. We discuss the impact of the drop \Pn on the overall mass transfer. In section~\ref{sec:conc}, we draw our conclusions.

\section{Mathematical model}\label{sec:mathmodel}

\subsection{Characteristic time scales}\label{sec:charatimescales}

Different time scales can be related to the main physical processes involved in the transport of a passive tracer from a viscous droplet (denoted with a subscript $d$) into a submerging falling film (denoted with a subscript $f$), as depicted schematically in figure~\ref{fig:droplet-film}. A brief study of the time scales corresponding to each process of the convective mass transfer can provide some insight for the modelling of the temporal evolution of the tracer concentration in the drop.  {Typical experimental values for the key parameters in this problem are summarized in table~\mbox{\ref{tab:expcharacpara}}. The experimental procedures describing how these parameters were chosen or measured are described in section~\mbox{\ref{sec:xpproc}}. The time scales calculated in this section and summarized in table~\mbox{\ref{tab:expcharactime}} are based on our experimental results.}

\begin{table}
	\scriptsize 
	\begin{minipage}{\textwidth}
		\centering
		\begin{tabular}{@{}cccccccc@{}}
			\hline \\
			Drop  &   Drop  &  Contact  &  Diffusion &  Dynamic  &  Inclination  &   Film   &  Film \\
			height &   length &   angle &  coefficient &   viscosity  &   angle &   kinematic &   height \\
			 &    &    &   &    ratio &    &    viscosity &   \\ \\  	\hline  \\
			$h_d$ (m) & $L$ (m) & $\theta_a$ (rad) & $D$ (m$^2\;$s$^{-1}$) & $\mu_f/\mu_d$ & $\alpha$ (rad) & $\nu_f$ (m$^2\;$s$^{-1}$) & $h$ (m) \\[5pt]
			
			
			 $3$--$6\times10^{-4}$ &  $1$--$4\times10^{-3}$ & $10^{-1}$--$1$ &$3$--$5\times10^{-10}$ &  $10^{-5}$ &  $3.5$--$7.9\times10^{-1}$ & $10^{-6}$ & $0.6$--$0.9\times10^{-3}$ \\ \\ \hline
			
		\end{tabular}
	\end{minipage}
	\caption{ {Experimental values for the main  parameters in the convective mass transfer problem from a droplet (subscript $d$) into a thin falling film (subscript $f$).}}
	\label{tab:expcharacpara}
\end{table}

At the interface separating the drop from the film flow we assume thermodynamic equilibrium and the mass transfer is considered instantaneous \cite[][]{faghri06}. According to \cite{koncsag11}, the interfacial mass transfer mechanism is well described by the stationary double film theory \cite[see also][]{bird07}, in the case of two quiescent phases. Although in our study the mass transfer in the bulk of the film phase, and possibly the drop phase in some parameter regimes, is strongly dominated by advection, reviewing the stationary double film theory can help us understand the physics of the mass transport in the vicinity of the interface. We consider the case of a mass transfer of a species $B$ from a quiescent liquid phase 1 into a quiescent liquid phase 2. Firstly, $B$ diffuses from the bulk of phase 1 towards the interface. Secondly, at thermodynamic equilibrium, the transfer across the interface from 1 into 2 is governed by the Nernst distribution law, which characterizes a jump in the concentration of $B$, such that
\begin{equation}\label{eq:jumpCint}
C_{B2i} = \beta C_{B1i},
\end{equation}
where $C_{B1i}$ and $C_{B2i}$  are the  equilibrium concentrations of $B$ at the interface in phase 1 and phase 2, respectively, and $\beta$  is a coefficient which typically depends on the temperature according to the Gibbs law \cite[][]{koncsag11}. \cite{lewis54} noted that the overall mass transfer could depend on the bulk concentration of $B$ in one or both phases, interfacial turbulence and any increase in the transfer area, or even the occurrence of chemical reactions at the interface. \cite{lewis54} further emphasized that the overall mass transfer is always limited by the slowest of the three consecutive steps described above: the diffusion from the bulk of phase 1 to the interface, the transfer across the interface, or the diffusion from the interface to the bulk of phase 2.

In the film phase, we assume that the thin  {drop (aspect ratio $\eta_d=h_d/L \approx0.1$)} does not affect strongly the flow in the film \cite[][]{baines:james94,danberg08,blount10}. The film flow is fully developed and follows a viscous--gravity regime known as flat Nusselt film regime \cite[\eg][]{kalliadasis12}. Applying the lubrication approximation, we can neglect all the velocity components except the streamwise component, such that the velocity field  is $\boldsymbol{u}=(u(y),0,0)$, with
\begin{equation}\label{eq:viscousgravU}
u = \frac{\gamma h}{2} \( 2 \frac{y}{h} - \(\frac{y}{h}\)^2 \),
\end{equation}
with $h$ the  film height, $y$ the coordinate in the direction normal to the substrate  with the origin $y=0$ at the substrate (see figure~\ref{fig:droplet-film}), and where $\gamma$, the film shear rate is
\begin{equation}
\gamma = \frac{g \sin \alpha h}{\nu_f},
\end{equation}
with $g$ the acceleration due to gravity, $\alpha$ the inclination angle of the substrate from the horizontal and $\nu_f$ the film kinematic viscosity.  {The typical value in the present study is $\gamma \approx \SI{e3}{s^{-1}}$.} The characteristic velocity  is
\begin{equation}
	U_f=\int_{0}^{h}\frac{u(y)}{h} \: \dd y=\frac{\gamma h}{3}, 
\end{equation}
the depth-averaged streamwise velocity in the film.  {The typical value in the present study is $U_f \approx \SI{1}{m. s^{-1}}$.} The film advective time scale, which corresponds to the characteristic time for the flow to pass the drop, is
\begin{equation}\label{eq:taufadv}
\tau_{f,adv} = \frac{L}{U_f}.
\end{equation}
  {The typical value in the present study is $\tau_{f,adv} \approx \SI{e-3}{s}$.} This is the shortest characteristic time in the problem. The film diffusion time scale, which corresponds to the characteristic time for the tracer to diffuse through the film thickness, is
\begin{equation}\label{eq:taufdif}
\tau_{f,dif} = \frac{h^2}{D_f},
\end{equation}
where the diffusivity $D_f$ of the tracer in the film phase is of the order of $10^{-10}\;$m$^2\;$s$^{-1}$ according to our experimental measurements (see section~\ref{sec:xpproc} and Appendix~\ref{apx:DiffExp}).  {The typical value in the present study is $\tau_{f,dif} \approx \SI{e3}{s}$.} The characteristic \Pn in the film is of the order of $\Pe_f = \tau_{f,dif}/\tau_{f,adv} \approx 10^6$. Therefore, transport in the film is strongly dominated by advection processes. As the tracer transfers from the drop into the flowing film, an advective--diffusive boundary layer forms above the drop, whose thickness can be expressed  as \cite[][]{baines:james94,blount10}
\begin{equation}\label{eq:deltafbl}
\delta = \(\frac{D_f L}{\gamma}\)^{1/3},
\end{equation}
where $L\approx 10^{-3}\:$m is the drop  {length. The typical value in the present study is $\delta \approx \SI{e-6}{}$ to $ \SI{e-5}{m}$. 
Effectively,} $\delta$ corresponds to the distance in the $y$-direction travelled by a passive tracer under the action of diffusion, in the time taken by the film flow to transport this tracer over the length of the drop.  {To compute $\delta$, we make the well-known L\'ev\^eque assumption that the velocity is linear in the diffusive boundary layer} \cite[][]{leveque28,blount10,bejan13},   {which is satisfied since $\delta \ll h$ by almost two orders of magnitude}. The characteristic transport time in the diffusive boundary layer is
\begin{equation}\label{eq:taudelta}
\tau_{\delta} = \frac{\delta^2}{D_f} = \frac{L}{\gamma \delta}.
\end{equation}
 {The typical value in the present study is $\tau_{\delta} \approx \SI{e-2}{}$ to $ \SI{e-1}{s}$.} Note that, by construction, the \Pn in the film diffusive boundary layer is $O(1)$.

In the drop phase, due to the shear imposed by the film at the drop--film interface, a recirculation flow can develop \cite[][]{honerkamp13,dimitrakopoulos98}. Neglecting surface tension and assuming the continuity of the tangential shear stress as well as the continuity of the tangential velocity at this liquid--liquid interface, the characteristic recirculation velocity in the drop phase is
\begin{equation}\label{eq:Ud}
U_d=\frac{\mu_f}{\mu_d} \gamma h_d,
\end{equation}
in the case of a thin drop.  {The typical value in the present study is $U_d \approx \SI{e-6}{}$ to $ \SI{e-5}{m. s^{-1}}$.} The viscosity ratio between the two phases is $\mu_f/\mu_d \approx 10^{-5}$, with $\mu_f=10^{-3}\;$Pa$\;$s the dynamic viscosity of water for the film phase, and $\mu_d \approx 10^{2}\;$Pa$\;$s the typical dynamic viscosity of the drop phase.  {In the present model, we assume a Newtonian rheology in the drop phase. We will return to this point in section~\mbox{\ref{sec:xpproc}}, when describing the particular rheology of the drop phase used in our experiments.} 
The characteristic height of the droplet is approximately $10^{-4} < h_d < 10^{-3}\;$m in the present study. 
Hence, the characteristic recirculation or advection time scale in the drop is
\begin{equation}\label{eq:taudadv}
\tau_{d,adv} = \frac{L}{U_d} = \frac{\mu_d}{\mu_f \gamma \eta_d}.
\end{equation}
 {The typical value in the present study is $\tau_{d,adv} \approx \SI{e2}{}$ to $ \SI{e3}{s}$.} \cite{dussan87} found theoretically that the characteristic velocity inside a drop scales like $U_d \propto \mu_f/\mu_d \gamma L \theta_a$, with $\theta_a \approx 0.1$ to $1\;$rad the advancing contact angle. We  note that in the limit of thin droplets, $\theta_a\approx 2h_d/L \ll 1$, the relationship of \cite{dussan87} is equivalent to our estimation for $U_d$ in (\ref{eq:Ud}).
In comparison, the characteristic diffusion time  in the drop is
\begin{equation}\label{eq:taudDif}
\tau_{d,dif}=\frac{\mbox{$h_d$}^2}{D_d}, 
\end{equation}
with the drop diffusivity $D_d\approx D_f\approx10^{-10}\;$m$^2\;$s$^{-1}$.  {The typical value in the present study is $\tau_{d,dif} \approx 10$ to $ \SI{e2}{s}$.} We can compute an effective \Pn characterizing the transport inside the drop:
\begin{equation}\label{eq:dropPn}
\Pe_{d}=\frac{\tau_{d,dif}}{\tau_{d,adv}}=\frac{\mu_f \gamma \eta_d \mbox{$h_d$}^2}{\mu_d D_d}. 
\end{equation}
 {The typical value in the present study is $\Pe_{d} \approx \SI{e-2}{}$ to $1$.} In the limit of low drop \Pn, $\Pe_{d} \ll 1$, we expect the transport of the tracer inside the drop to be dominated by diffusion. At $\Pe_{d} \approx 1$, we believe that the transport within the drop is a combination of both advection and diffusion processes. We  note that $\tau_{d,adv}$ is based on the drop length $L$ as characteristic length, whereas $\tau_{d,dif}$ is based on the drop thickness $h_d$. Hence, the drop aspect ratio $\eta_d=h_d/L$ is also an important physical parameter for the mass transport in the drop. In this model, we consider the case of thin droplets with $\eta_d \approx 0.1$.

Another time scale can be associated with the drop: its deformation time scale $\tau_{d,def}$. The drop elongates, mainly in the streamwise direction, due to the shear stress exerted by the film flow on its surface \cite[\eg][]{puthenveettil13,ahmed13}. From our experimental measurements, we can estimate the characteristic time of the drop elongation based on an empirical exponential model (see equation (\ref{eq:Aint}) in Appendix~\ref{apx:Aint}): $\tau_{d,def} \approx 1$ to $\SI{10}{s}$. 
In our experiments, the deformation time scale is smaller than the drop displacement time scale associated with the motion of the drop centre of mass. The time scale for the centre of mass of a drop to be displaced by a distance equal to the drop initial length $L$ is $\tau_{d,dis} \approx 10$ to $100\;$s, according to our measurements.

We have summarized in table~\ref{tab:expcharactime}  the typical values for the present study of all the characteristic time scales  involved in the convective mass transfer from a viscous drop into a submerging falling film. As we can see, this problem spans a broad range of time scales: from $\tau_{f,adv} \approx \SI{e-3}{s}$ for the shortest advection time scale in the film to $\tau_{f,dif} \approx \SI{e3}{s}$ for the longest diffusive time scale in the film.

\begin{table}
	\begin{minipage}{\textwidth}
		\centering
		\begin{tabular}{@{}lcll@{}}
			\hline \\
			$\tau_{f,adv}$  \hspace{0cm} & $\sim$ \hspace{0cm} & $10^{-3}\;$s & \hspace{0cm} $x$-advection in the film bulk along the drop length $L$ (see (\ref{eq:taufadv})) \\
			$\tau_{f,dif}$ \hspace{0cm} & $\sim$ \hspace{0cm} & $10^{3}\;$s & \hspace{0cm} $y$-diffusion in the film bulk across the film height $h$ (see (\ref{eq:taufdif})) \\
			$\tau_{\delta}$ \hspace{0cm} & $\sim$ \hspace{0cm} & $10^{-2}$--$10^{-1}\;$s & \hspace{0cm} transport in the film diffusive boundary layer (see (\ref{eq:taudelta})) \\
			$\tau_{d,adv}$ \hspace{0cm} & $\sim$ \hspace{0cm} & $10^2$--$10^{3}\;$s & \hspace{0cm} $x$-advection in the drop over $L$ (see (\ref{eq:taudadv})) \\
			$\tau_{d,dif}$ \hspace{0cm} & $\sim$ \hspace{0cm} & $10$--$10^{2}\;$s & \hspace{0cm} $y$-diffusion in the drop bulk across the drop height $h_d$ (see (\ref{eq:taudDif})) \\
			$\tau_{d,def}$ \hspace{0cm} & $\sim$ \hspace{0cm} & $1$--$10\;$s & \hspace{0cm} characteristic drop deformation \\
			$\tau_{d,dis}$ \hspace{0cm} & $\sim$ \hspace{0cm} & $10$--$10^{2}\;$s & \hspace{0cm} displacement of the drop centre of mass over $L$\\
			$\tau_{\kappa}$ \hspace{0cm} & $\sim$ \hspace{0cm} & $10$--$10^{2}\;$s & \hspace{0cm}  transport from the bulk of the drop to the bulk of the film (see (\ref{eq:kappadef})) \\ 
 \\	\hline
			
		\end{tabular}
	\end{minipage}
	\caption{ {Typical values of the main characteristic time scales and associated physical processes in the convective mass transfer problem from a droplet into a thin falling film.}}
	\label{tab:expcharactime}
\end{table}

\subsection{Overall mass transfer}\label{sec:masstransModel}

To model the overall convective mass transfer related to the change of tracer concentration in the drop in time, we use an analogy with Newton's empirical law of cooling \cite[see \eg][]{kays05}. The change in time of the total mass of tracer in the droplet is proportional to the concentration difference between the  spatially averaged concentration in the drop $C_d$ and the fixed background concentration in the  environment $C_{\infty}$,
\begin{equation}\label{eq:NewtonCooling}
F = \frac{\dd V C_d}{\dd t} = -\kappa A \(C_d-C_{\infty}\),
\end{equation}
where the concentrations $C_d$ and $C_{\infty}$ are volumetric masses of tracer measured in units of density, $V$ is the total volume of the drop, $t$ is time, $\kappa$ is a convective mass transfer coefficient expressed as a velocity, and $A$ is the drop--film interface area. We consider the case where the quantity of tracer in the drop is small, $C_d/\rho_d \ll 1$ (with $\rho_d$ the bulk density of the drop). We assume that the total volume of the drop remains approximately constant in time such that $V=V_0$. Moreover, we assume that both $A$ and $\kappa$ remain constant in time. Therefore, the solution of (\ref{eq:NewtonCooling})  is
\begin{equation}\label{eq:Csolution}
\hat{C}_d = e^{-\hat{t}},
\end{equation}
where we normalised the drop concentration and time such that
\refstepcounter{equation}
$$
  \hat{C}_d = \frac{C_d-C_{\infty}}{ C_0 -C_{\infty}}, \quad \hat{t} = \frac{t}{\tau_{\kappa}},
  \eqno{(\theequation{\textrm{\textit{a,b}}})}\label{eq:normCt}
$$
in which hats denote non-dimensional quantities, $C_0$ is the initial uniform concentration in the droplet and
\begin{equation}\label{eq:kappadef}
\tau_{\kappa} = \frac{V_0}{A\kappa},
\end{equation}
the characteristic time scale for the evolution of the concentration in the drop. According to our experimental results, presented in section~\ref{sec:xpresults}, we have $\tau_{\kappa} \approx 10$ to $100 \;$s. As mentioned previously, in practice, the drop can deform and elongate slightly. This deformation induces an increase in the interface area $A$, empirically modelled in (\ref{eq:Aint}) in Appendix~\ref{apx:Aint}, which we can fit with our experimental data. However, we  note from table~\ref{tab:expcharactime}, that in our experiments the deformation time scale $\tau_{d,def}\approx 1$ to $10\;$s is smaller than the characteristic overall mass transfer time scale $\tau_{\kappa}\approx 10$ to $100\;$s. Therefore, the drop rapidly adjusts to its final shape with $A \rightarrow A_m$ (the maximum interface area) as $t>\tau_{d,def}$. Note that the displacement time scale, which accounts for a solid-body type of displacement of the drop, should not influence the mass transfer because $\tau_{d,dis}\gg \tau_{\delta}$. Transport processes in the diffusive boundary layer above the interface occur much faster than the deformation and displacement of the interface  {since $\mu_d \gg \mu_f$}. The  shape of the drop can effectively be considered as steady and having reached its final size and shape for the convective mass transfer. We also solve in  {\mbox{Appendix~\ref{apx:Aint}} (equation \mbox{(\ref{eq:Csolution2})}) the case where the change in time of the drop interface area} can affect the mass transfer. This case is observed for lower drop viscosity, with the film shear forces elongating the drop by a significant amount, up to several times the initial drop length \cite[][]{puthenveettil13}.

The time scale $\tau_{\kappa}$ in (\ref{eq:kappadef}) characterizes the total transport time taken for the tracer to migrate from the bulk of the drop to the bulk of the film flow.  {This time scale is effectively the sum of the characteristic transport times through each phases, since the transport through each phase occurs consecutively}. Assuming that the transfer time across the interface is negligible and  the interface cannot store any tracer, we have
\begin{equation}
\tau_{\kappa}\approx \tau_d+\tau_f,
\end{equation}
where  {$\tau_d$ is the characteristic transport time in the drop phase, and $\tau_f$ is the characteristic transport time in the film phase, above the drop}. In general, $\tau_d$ and $\tau_f$ can depend on a combination of both advection and diffusion processes, or only the fastest of these two processes. In the present study, the transport time in the bulk of the drop $\tau_d$ can depend on both advection and diffusion processes because the drop \Pn ranges  {$Pe_d \approx 10^{-2}$} to $1$. The characteristic transport time can thus be defined as $\tau_d \propto \min(\tau_{d,dif},\tau_{d,adv})$ (with $\min(\cdot,\cdot)$ the function taking the minimum of its arguments). On the other hand, advection processes strongly dominate in the bulk of the film phase since  $\Pe_f \approx 10^6$. In fact the transport time in the film $\tau_f$ corresponds mainly to the transport time through the diffusive boundary layer because $\tau_{\delta}\gg \tau_{f,adv}$. We have $\tau_f \approx \tau_{\delta}$, which expresses a balance between advection and diffusion in the thin diffusive boundary layer in the film phase. Furthermore, according to table~\ref{tab:expcharactime}, we have  $\tau_d \gg \tau_{\delta}$ for all the regimes investigated. Therefore, in the present study, the total transport time $\tau_{\kappa}$ depends mainly on transport inside the drop. We can distinguish two regimes for $\tau_{\kappa}$, depending on the drop \Pn:
\begin{eqnarray}
\tau_{\kappa} & \approx & \tau_{d,dif}, \  \Pe_d\ll 1, \quad \textrm{the drop diffusion-dominated regime (D)}; \label{eq:regimeD} \\
\tau_{\kappa} & \approx & \tau_{d,dif} \approx \tau_{d,adv}, \  \Pe_d\approx 1, \quad \textrm{the drop advection--diffusion regime (AD)}. \label{eq:regimeAD}
\end{eqnarray}
 {A} third regime, where advection dominates transport in the bulk of the drop, could exist at large drop \Pns. In our study we have $Pe_d \leq 1$, therefore we only investigate the regime where diffusion is dominant in the drop (regime D), and the regime where both advection and diffusion are important in the drop (regime AD).

\subsection{Mass transport in the film phase}\label{sec:transfilm}

The only unknown in the Newton-cooling model (\ref{eq:NewtonCooling}) is the mass transfer coefficient $\kappa$. In this and the two following sections, one of the main objectives is to determine theoretically $\kappa$, or its non-dimensional form the Sherwood number $\Sh=\kappa \ell/D$, by modelling the mass transfer in both the film phase and the drop phase, as well as the coupling at the interface. Another objective is to support the  empirical Newton-cooling model through theoretical modelling from fundamental physical principles. In this section, we focus on the transport in the thin diffusive boundary layer developing above the drop--film interface, in the film phase.

Following the works of \cite{baines:james94} and \cite{blount10}, we solve analytically the two-dimensional, steady advection--diffusion equation for the spatial evolution of the tracer concentration in the diffusive boundary layer
\begin{equation}\label{eq:advdiff1}
\boldsymbol{u} \bcdot \bnabla C  = D_f \bnabla^2 C,
\end{equation}
where $\boldsymbol{u}$ is the film velocity field, and $\bnabla$ is the gradient operator, both along the streamwise ($x$) and normal directions ($y$). We use the  boundary conditions
\refstepcounter{equation}
$$
  C = C_{f,i},\ y = h_d, \ 0 < x < l_d \ \textrm{and} \ C \rightarrow C_{\infty},\ y-h_d \gg \delta,
  \eqno{(\theequation{\textrm{\textit{a,b}}})}\label{eq:BCadvdifffilm}
$$
with $C_{f,i}$ the concentration just above the drop--film interface in the film phase and $0 \leq l_d \leq L$ the streamwise length of the drop at a particular transverse location $z$ over the drop. We can neglect the time dependence of the concentration in (\ref{eq:advdiff1}) because the characteristic time scale for the establishment of the diffusive boundary layer, $\tau_{\delta}\approx 10^{-2}$ to $10^{-1}\;$s, is much smaller than the characteristic time scale for the evolution of the drop concentration, $\tau_{\kappa}\approx 10$ to $100\;$s. Effectively, we assume that the interfacial concentration $C_{f,i}$ is quasi-steady for the mass transport in the diffusive boundary layer. The unknown interfacial concentration $C_{f,i}$ is determined by coupling the transport in the film phase with the transport in the drop phase, as will be shown in section~\ref{sec:modeldropphase}. We also discuss further in section~\ref{sec:modeldropphase} the impact of the slow time-dependence of $C_{f,i}$. In (\ref{eq:advdiff1}) we assume  that the film flow is laminar, fully developed, and the velocity field is semi-parabolic following the viscous--gravity regime described in (\ref{eq:viscousgravU}). At the drop--film interface, due to the continuity of the tangential shear stress and the tangential velocity, the fluid has a small interfacial velocity $U_i$.  {The interfacial velocity $U_i$ is of the order of magnitude of the drop characteristic velocity scale $U_d\approx 10^{-6}$ to $10^{-5}$\mpers estimated from dimensional analysis in \mbox{(\ref{eq:Ud})} \mbox{\cite[see also][for further details]{dussan87}}}. We  note that $U_i$ is much smaller than the characteristic velocity in the diffusive boundary layer  {$U_{\delta}=L/\tau_{\delta}\approx 10^{-2}$ to $10^{-1}$}\mpers. Thus, we can neglect the influence of the interfacial velocity $U_i$.

 {Since the diffusive boundary layer thickness, $\delta \approx \SI{e-6}{}$ to $ \SI{e-5}{m}$, is much smaller than the film thickness, $h \approx \SI{e-4}{}$ to $ \SI{e-3}{m}$,} we can use L\'ev\^eque's (1928) assumption: $u\approx \gamma y$ in the diffusive boundary layer. We  assume that the characteristic film shear $\gamma$ is, in first-order approximation, unperturbed by the presence of the drop.  {We do not take into account a varying $\gamma$ and do not consider flow separation.
We} also assume that diffusion processes in the streamwise direction are negligible compared with advection processes. Thus, (\ref{eq:advdiff1}) becomes
\begin{equation}\label{eq:advdiff2}
3U_f \frac{\(y-h_d\)}{h} \frac{\p C}{\p x}  = D_f \frac{\p^2 C}{\p y^2} \quad \textrm{for} \ y>h_d \ \textrm{and} \ 0 < x < l_d.
\end{equation}
Contrary to the problems studied by \cite{stone89}, \cite{baines:james94}, and \cite{danberg08}, who considered a general shear flow in a semi-infinite environment, here the flow has a well-defined characteristic velocity $U_f=\gamma h/3$ and a well-defined characteristic length scale $h$. We can use the following non-dimensionalisation, which is similar to the method used by \cite{bejan13} for the case of heat transfer in a fully-developed pipe flow (where the pipe centreline is analogous to the free surface in our problem),
\refstepcounter{equation}
$$
  \breve{C} = \frac{C-C_{\infty}}{C_{f,i}-C_{\infty}}, \quad \breve{y} = \frac{y-h_d}{h}, \quad \breve{x} = \frac{x}{h \Rey_f \Sc_f},
  \eqno{(\theequation{\textrm{\textit{a--c}}})}\label{eq:ndCfilm}
$$
where breves denote non-dimensional variables, and with the film Reynolds number $\Rey_f=h U_f/\nu_f$ and the film Schmidt number $\Sc_f=\nu_f/D_f$. 
We also assume in (\ref{eq:advdiff2}) that the mass flux of tracer is very small compared with the characteristic mass flux of water in the film. As \cite{kays05} explain, (\ref{eq:advdiff2}) is valid for negligible normal velocity at the interface. At the drop--film interface, the thermodynamic equilibrium is maintained if the tracer transferring into the film flow do not affect the chemical potential, the temperature (through the heat of dilution), or  the viscosity \cite[][]{faghri06}. Since $C \ll \rho_f$ ( {where $C$ is expressed in units of density, as mentioned before,} and with $\rho_f$ the bulk density of the film phase) and the flux of tracer from the drop into the film is mainly driven by diffusion, we expect that the assumptions above are satisfied and the  diffusive boundary layer in the film is not disrupted by the mass transfer.
Equation (\ref{eq:advdiff2}), under the boundary conditions (\ref{eq:BCadvdifffilm}\textit{a,b}), using the non-dimensionalisation defined in (\ref{eq:ndCfilm}\textit{a--c}), can be solved with a similarity form using the similarity variable $\breve{y}/\breve{x}^{1/3}$ \cite[][]{baines:james94,blount10}. We find
\begin{equation}\label{eq:solC}
\breve{C}  = \frac{1}{\Gamma\[1/3\]} \Gamma\[\frac{1}{3},\frac{\breve{y}^3}{3\breve{x}}\],
\end{equation}
where $\Gamma[g]=\int_{0}^{\infty} s^{g-1} e^{-s} \: \mathrm{d}s$ is the Gamma function and $\Gamma[g,\zeta]=\int_{\zeta}^{\infty} s^{g-1} e^{-s} \: \mathrm{d}s$ is the upper incomplete Gamma function. The mass flux per unit area at the drop--film interface is defined as
\begin{equation}
j  = -\frac{D_f}{h} \(C_{f,i}-C_{\infty}\) \frac{\p \breve{C}}{\p \breve{y}}\(\breve{y}=0 \),
\end{equation}
with the convention $j>0$ for a positive flux from the drop into the film. We obtain
\begin{equation}\label{eq:massfluxj}
j  = \frac{D_f}{h} \(C_{f,i}-C_{\infty}\) \frac{3^{2/3}}{\Gamma\[1/3\]} \(\frac{x}{hRe_f Sc_f}\)^{-1/3}.
\end{equation}
We can compute the total mass flux for a drop by integrating  {$j$} over the area $A$ of the interface
\begin{equation}\label{eq:fluxFf}
F_f  = \int\!\!\!\int_{A} j \dd A.
\end{equation}
We assume that the drop--film interface is a spherical cap of base diameter $L$ and maximum height $h_d$. The flux is integrated first along the streamwise direction for $0<x<l_d(z)$, as we assumed a two-dimensional flow in the $(x,y)$ plane. Subsequently, we can integrate the flux in the cross-stream direction for $-L/2<z<L/2$. Since the drop aspect ratio $\eta_d=h_d/L \approx 0.1$ is very small, we can use a Taylor expansion for $\eta_d\ll 1$. We find the total integrated flux over the drop spherical cap
\begin{equation}
F_f  = \frac{3^{2/3}\sqrt{\pi}}{\Gamma\[5/6\]} \(\(\frac{L}{h}\)^2 \Rey_f \Sc_f\)^{1/3} D_f \(C_{f,i}-C_{\infty}\) L \(\frac{3}{10}+\frac{172}{165}\mbox{$\eta_d$}^2 + O\(\mbox{$\eta_d$}^4\) \).
\end{equation}
We can then compute the Sherwood number for the convective mass transfer from the drop into the film: it is generally defined as $\Sh=\kappa \ell/D$, with $\kappa$ the convective mass transfer coefficient and $\ell$ a characteristic length scale. In our problem, the spatially averaged Sherwood number for the drop is
\begin{equation}\label{eq:defSh}
\overline{\Sh_f} = \frac{F_f \ell}{D_f \(C_{f,i}-C_{\infty}\) \(\int\!\!\!\int_{A} \dd A\)},
\end{equation}
with
\begin{equation}
\ell=\(\int\!\!\!\int_{A} \dd A\)^{1/2},
\end{equation}
 {chosen as the characteristic length for the rest of this study}. Consequently, we obtain
\begin{equation}\label{eq:SherNum2order}
\overline{\Sh_f} =  \frac{3^{2/3}}{\Gamma\[5/6\]} \(\frac{3}{5}+\frac{146}{165}\mbox{$\eta_d$}^2 \) \(\(\frac{L}{h}\)^2 \Rey_f \Sc_f\)^{1/3},
\end{equation}
at the third order in $\eta_d$. Since the correction due to the drop aspect ratio is very small (second order), we can compute an approximation at the first order in $\eta_d$ as
\begin{equation}\label{eq:SherNum1order}
\overline{\Sh_f} =  0.766 \mbox{$\Rey_L$}^{1/3} \mbox{$\Sc_f$}^{1/3},
\end{equation}
where the characteristic Reynolds number is
\begin{equation}
\Rey_L = \frac{\gamma L^2}{\nu_f}.
\end{equation}
 {We  note that, although we initially used the film thickness $h$ as characteristic length scale for the film flow (see \mbox{(\ref{eq:viscousgravU})}), the drop length $L$ emerges as the natural length scale in \mbox{(\ref{eq:SherNum1order})}. This result is consistent with other studies} \cite[][]{stone89,baines:james94,danberg08},  {in which $L$ was a priori selected as the characteristic length scale. In these studies, $L$ was the only characteristic length scale as the domain was infinite and  the drop thickness was assumed negligible. In the mass transfer problem described in figure~\mbox{\ref{fig:droplet-film}}, we can perceive that the only macroscopic length scale which is `seen' by the two-dimensional diffusive boundary layer at the surface of the drop is the distance from the upstream edge of the drop. The mass transfer is not, at least directly, impacted by the thickness of the drop $h_d$. Moreover, it is not affected by the film thickness $h$, unless the diffusive boundary layer grows sufficiently large, which is very unlikely in the regimes considered in the present study where $\delta/h \approx 10^{-3}$ to $10^{-2}$}  \cite[see][for the case where the diffusive boundary layer thickness is not small compared with the thickness of the bulk flow]{bejan13}.  {We  discuss further the physical interpretation of \mbox{(\ref{eq:SherNum1order})} and $\Rey_L = \gamma L^2/\nu_f$ in \mbox{section~\ref{sec:mtFilm}}, where we  propose an alternative to \mbox{(\ref{eq:SherNum1order})}, mathematically equivalent but with a more intuitive physical interpretation.}

We also note that the Sherwood number $\overline{\Sh_f}$, as defined in (\ref{eq:defSh}), cannot be fully computed at this point because this definition depends on the unknown interfacial concentration $C_{f,i}$. We shall resolve this situation in section~\ref{sec:sherwoodnum} using the spatially averaged concentration in the drop determined in the next section.

\subsection{Mass transport in the drop phase}\label{sec:modeldropphase}

To support our empirical Newton-cooling model, we now model transport inside the drop based on fundamental physical principles. This new model describes mainly the regime D explained in section \ref{sec:masstransModel}, where for small drop \Pns transport  in the drop is dominated by diffusion. We believe this model is important in the case considered in the present study as the overall mass transfer time $\tau_{\kappa}$ should scale like the characteristic diffusive time  in the drop $\tau_{d,dif}$ for $\Pe_{d}\ll 1$. We couple this model of transport within the drop with the model for the transport in the film flow presented in the previous section through the boundary conditions at the drop--film interface.

For  {$\tau_f\ll \tau_d$}, the tracer concentration inside the drop is not uniform: a concentration gradient establishes in time between the interface and the bulk of the drop. The temporal evolution of the concentration in the drop, when submerged in a falling liquid film, can be modelled using a time-dependent diffusion equation. Advection processes in the drop are deemed negligible since we consider the case where  $\Pe_{d}\ll 1$. We also assume that lateral diffusion processes in the $x$ and $z$ directions are negligible compared with diffusion processes in the $y$ direction. We expect that this assumption is true for thin, flat droplets of fairly uniform thickness, except near the edges where three-dimensional effects can be important. Under these assumptions, we model the concentration in the drop $C(y,t)$ using a one-dimensional time-dependent diffusion equation through the height of the drop for $0\leq y < h_d$ and for $t>0$
\begin{equation}\label{eq:1ddiff}
	\frac{\p C}{\p t} = D_d \frac{\p^2 C}{\p y^2},
\end{equation}
where $D_d$ is the diffusion coefficient inside the drop. We use the  boundary and initial conditions
\refstepcounter{equation}
$$
C = C_{d,i}(t),\ y = h_d, \ t > 0; \ \frac{\p C}{\p y} = 0,\ y=0, \ t\geq 0;  \ \textrm{and} \ C = C_0, \ 0 \leq y < h_d, \ t=0;
\eqno{(\theequation{\textrm{\textit{a--c}}})}\label{eq:1ddiffBC}
$$
with $C_{d,i}$  {the time-dependent concentration} inside the drop just below the interface and $C_0$ the initial uniform concentration in the  {drop. 
Due to the small concentration of tracer in the drop, the problem of mass diffusion described above is effectively analogous to the conduction of heat in a one-dimensional solid, insulated at one end and with a time-dependent temperature at the other end \mbox{\cite[\eg][]{incropera07}}.}

We can couple the transport inside the drop with the transport in the film by assuming the continuity of both the concentration (defined as mass density) and the flux at the drop--film interface
\refstepcounter{equation}
$$
C_{d,i} = C_{f,i}=f(t),\ y = h_d, \ t > 0; \ \frac{\p C_{d,i}}{\p y} = \xi \overline{\frac{\p C_{f,i}}{\p y}},\ y=h_d, \ t > 0,
\eqno{(\theequation{\textrm{\textit{a--b}}})}\label{eq:1ddiffBC2}
$$
where $\xi=D_f/D_d$. The  function $f(t)$ represents the time-dependent concentration at the drop--film interface. We replace the boundary condition (\ref{eq:1ddiffBC}\textit{a}) in the initial boundary value problem with (\ref{eq:1ddiffBC2}\textit{a}). The unknown function $f(t)$ will be found using the equation stating the continuity of the flux at the interface (\ref{eq:1ddiffBC2}\textit{b}). As mentioned in section \ref{sec:transfilm}, determining $f(t)$ is also important to compute the Sherwood number $\overline{\Sh_f}$ defined in (\ref{eq:defSh}). As explained previously, the slow time-dependent evolution of the concentration in the drop impacts the quasi-steady transport in the film boundary layer only through the change in time of the concentration at the interface, as expressed in   {\mbox{(\ref{eq:1ddiffBC2}\textit{a})}. 
The} overbar in (\ref{eq:1ddiffBC2}\textit{b}) represents a spatial average over the drop--film interface. According to (\ref{eq:massfluxj}), the local  flux $\p C_{f,i}/\p y$ at the drop--film interface depends on the distance from the upstream edge of the drop $x$. In order to simplify the model for the transport in the drop, which is based on a one- {dimensional  equation}, we assume a uniform mass flux over the drop interface and use the spatially averaged mass flux (see (\ref{eq:defSh})) from the model for the transport in the film, such that
\begin{equation}\label{eq:flux_interface}
D_f \overline{\frac{\p C_{f,i}}{\p y}} = \frac{F_f}{\int\!\!\!\int_{A} \dd A}.
\end{equation}

We non-dimensionalise the variables in the initial boundary value problem (\ref{eq:1ddiff}--\ref{eq:1ddiffBC2}) using
\refstepcounter{equation}
$$
\tilde{C} = \frac{C-C_{\infty}}{C_0-C_{\infty}}, \quad \tilde{t} = \frac{t}{\tau_{d,dif}}, \quad \tilde{y} = \frac{y-h_d}{h_d},
\eqno{(\theequation{\textrm{\textit{a--c}}})}\label{eq:normCt2}
$$
in which tildes denote non-dimensional quantities and where $\tau_{d,dif}=\mbox{$h_d$}^2/D_d$ (see \ref{eq:taudDif}). Note that the non-dimensionalisation for the model of the transport in the film flow in (\ref{eq:ndCfilm}) is slightly different. Thus, the initial boundary value problem becomes  for $-1 \leq \tilde{y} < 0$ and for $\tilde{t}>0$
\begin{equation}\label{eq:1ddiffND}
\frac{\p \tilde{C}}{\p \tilde{t}} = \frac{\p^2 \tilde{C}}{\p \tilde{y}^2},
\end{equation}
with
\refstepcounter{equation}
$$
\tilde{C}=\tilde{f}(\tilde{t}),\ \tilde{y} = 0, \ \tilde{t} > 0; \ \frac{\p \tilde{C}}{\p \tilde{y}} = 0,\ \tilde{y}=-1, \ \tilde{t} \geq 0;  \ \textrm{and} \ \tilde{C} = 1, \ -1 \leq \tilde{y} < 0, \ \tilde{t}=0.
\eqno{(\theequation{\textrm{\textit{a--c}}})}\label{eq:1ddiffBC3}
$$
We can solve this partial differential equation with inhomogeneous time-dependent boundary conditions using the methods of separation of variables and eigenfunction expansion  {\mbox{\cite[\eg][]{strauss08}}}. We find
\begin{equation}\label{eq:1ddiffNDsol}
\tilde{C}(\tilde{y},\tilde{t}) = \tilde{f}(\tilde{t}) + \sum_{n=0}^{\infty}\frac{2}{\lambda_n}\(\tilde{f}(0)-1+\int_{0}^{\tilde{t}} \tilde{f}'(\tilde{\tau}) e^{\mbox{\scriptsize $\lambda_n$}^2\tilde{\tau}} \dd \tilde{\tau} \) e^{-\mbox{\scriptsize $\lambda_n$}^2\tilde{t}} \sin\(\lambda_n \tilde{y}\),
\end{equation}
with $\lambda_n=\pi(2n+1)/2$ for all integers $n\geq 0$. In order to find the function $\tilde{f}$ using the continuity of the flux (\ref{eq:1ddiffBC2}\textit{b}), we assume that $\tilde{f}'(\tilde{\tau})=-\delta(\tilde{\tau})$ in the integral in (\ref{eq:1ddiffNDsol}), where $\delta(\tilde{\tau})$ is the Dirac delta function. In other words, we assume that the concentration at the drop--film interface decreases very rapidly in time at the start of the experiment. Then, the rate of  change of $C_{d,i}$ in time is negligible for larger $\tilde{t}$. This non-physical infinite decrease at $t=0$ is due to the non-physical concentration distribution in our problem where the concentration jumps from $C_{d,i}=C_0>0$ just below the interface to $C_{f,i}=0$ just above the interface at $t=0$. In reality, the concentration discontinuity disappears very quickly as soon as the film flows over the drop. Diffusion acts very rapidly at the interface to smooth the concentration distribution across the interface. We believe that this assumption is appropriate provided that $\tilde{t}$ is not too small. Using the spatially averaged flux (\ref{eq:flux_interface}) and the boundary condition of continuity of the flux (\ref{eq:1ddiffBC2}\textit{b}), we thus obtain
\begin{equation}\label{eq:checkf}
\tilde{f}(\tilde{t}) \approx \frac{2\(2-\tilde{f}(0)\)\ell}{ 0.766 \mbox{$\Rey_L$}^{1/3} \mbox{$\Sc_f$}^{1/3} h_d \xi} \sum_{n=0}^{\infty} e^{-\mbox{\scriptsize $\lambda_n$}^2\tilde{t}}.
\end{equation}
The infinite series in both the function $\tilde{f}(\tilde{t})$ and its derivative $\tilde{f}'(\tilde{t})$ are convergent for $\tilde{t}>0$ (but not for $\tilde{t}=0$, where they are not defined). We can also show that $\tilde{f}'(\tilde{t})$  calculated using (\ref{eq:checkf}) is  {approximately} similar to $-\delta(\tilde{\tau})$, as assumed previously. Since our model cannot predict the initial concentration at the interface, we define arbitrarily its value as $\tilde{f}(0)=1$, which preserves continuity within the drop phase. Therefore, the concentration distribution across the drop height evolves in time following
\begin{equation}\label{eq:1ddiffNDsolsimp}
\tilde{C}(\tilde{y},\tilde{t}) \approx  \sum_{n=0}^{\infty} \(\frac{2 \ell}{ 0.766 \mbox{$\Rey_L$}^{1/3} \mbox{$\Sc_f$}^{1/3} h_d \xi} -  \frac{2 \sin\(\lambda_n \tilde{y}\)}{\lambda_n} \)  e^{-\mbox{\scriptsize $\lambda_n$}^2\tilde{t}} \quad \textrm{with} \ \lambda_n = \frac{\pi}{2}(2n+1).
\end{equation}
We can also compute the spatially averaged concentration inside the drop
\begin{equation}\label{eq:1ddiffNDsolavg}
\tilde{C}_d(\tilde{t}) \approx  \sum_{n=0}^{\infty} \( \frac{2 \ell}{ 0.766 \mbox{$\Rey_L$}^{1/3} \mbox{$\Sc_f$}^{1/3} h_d \xi} + \frac{2}{\mbox{ $\lambda_n$}^2} \) e^{-\mbox{\scriptsize $\lambda_n$}^2\tilde{t}}  \quad \textrm{with} \ \lambda_n = \frac{\pi}{2}(2n+1).
\end{equation}

\begin{figure}
	\centering
	\input{./DropCvsy-model}
	\includegraphics[width=0.48\textwidth]{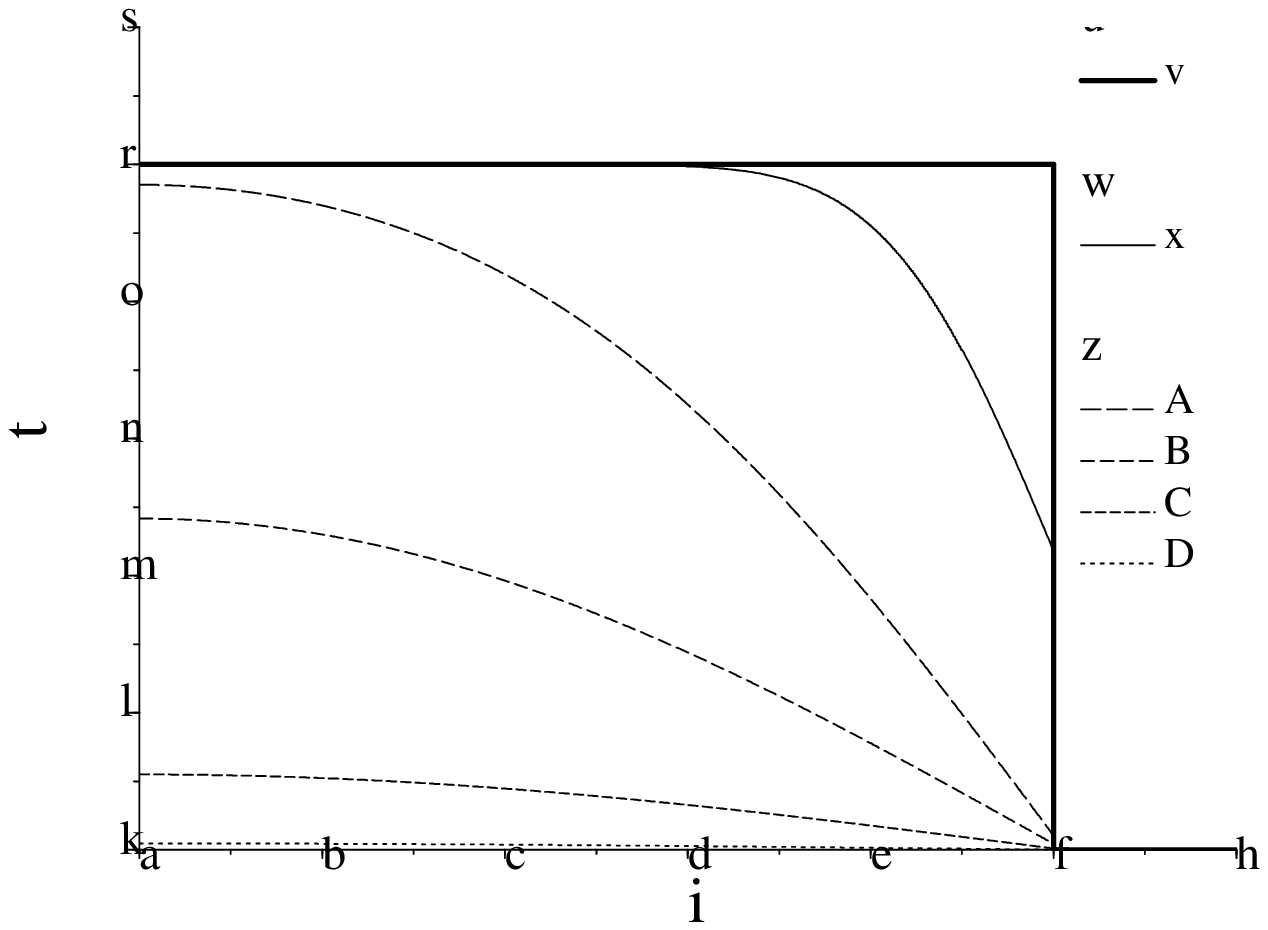}~~~~~
	\input{./AvgCvsT-model}
	\includegraphics[width=0.44\textwidth]{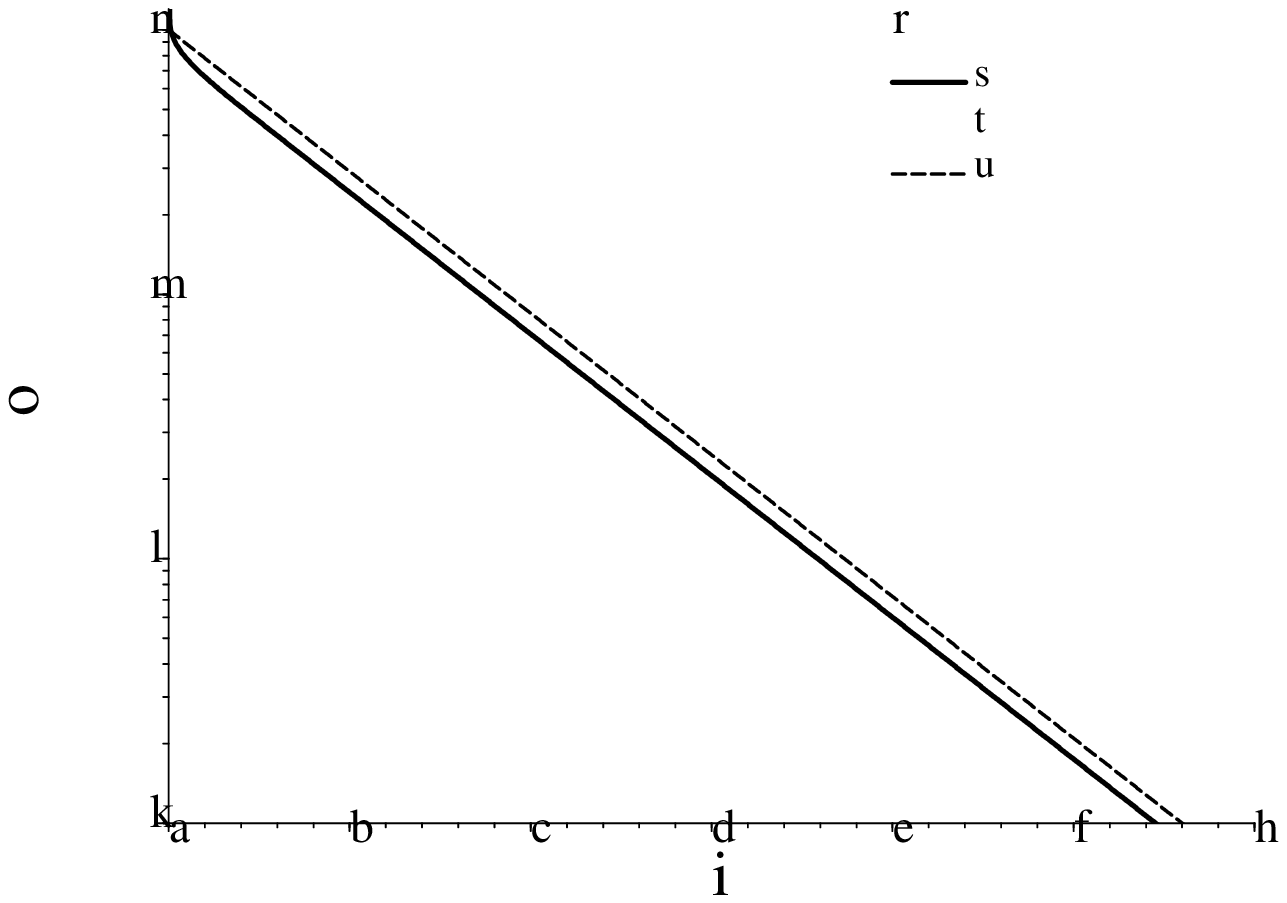}
	\caption{(\textit{a}) Distribution of the concentration at different times (plotted using different line styles) across the drop height, $-1 \leq \tilde{y} \leq 0$. The concentration $\tilde{C}$ is computed using (\ref{eq:purediffSolD}) for $\tilde{t}=0.01$ and using  (\ref{eq:1ddiffNDsolsimp}) for $\tilde{t}\geq 0.1$. (\textit{b}) Evolution of the spatially averaged drop concentration in time. We plot using a solid line on a logarithmic axis the concentration $\tilde{C}_d$ following (\ref{eq:1ddiffNDsolavg}). We also plot using a dashed line a simple exponential decrease in time. To compute both graphs we use typical values for the different parameters: $\Rey_L=30\thinspace 000$, $\Sc_f=2000$, $\xi=5/3$, $\ell=\sqrt{\pi} L/2$ and $\eta_d=h_d/L=0.15$.
	}
	\label{fig:DropCmodel}
\end{figure}

At small time where the assumption $\tilde{f}'(\tilde{\tau}) \approx -\delta(\tilde{\tau})$ is not valid, transport on both  {sides} of the drop--film interface is driven mainly by diffusion.  {This  diffusive regime} occurs for $t\leq \tau_{\delta}$, the establishment time of the diffusive boundary layer in the film.  {According to our experimental results, we have $\tilde{t}_{dif}=\tau_{\delta}/\tau_{d,dif}\approx 10^{-3}$ to $10^{-2}$. For $\tilde{t}\leq \tilde{t}_{dif}$, the characteristic diffusive length scale in the drop  $\sqrt{t_{dif} D_d}\approx 10^{-7}$ to $10^{-6}\;$m} is much smaller than the characteristic thickness of the droplet $h_d\approx 10^{-4}\;$m. Therefore, for $0<\tilde{t}\leq \tilde{t}_{dif}$, transport in the drop and the film phases can be considered as in  infinite domains, such that
\refstepcounter{equation}
$$
\frac{\p \tilde{C}}{\p t} = D_d\frac{\p^2 \tilde{C}}{\p y^2},\ \tilde{y}<0, \quad \textrm{and} \quad \frac{\p \tilde{C}}{\p t} = D_f\frac{\p^2 \tilde{C}}{\p y^2},\ \tilde{y}>0,
\eqno{(\theequation{\textrm{\textit{a,b}}})}\label{eq:purediff}
$$
respectively, with initial conditions at $\tilde{t} = 0$ satisfying
\refstepcounter{equation}
$$
\tilde{C}=1,\ \tilde{y} < 0, \quad \textrm{and} \quad \tilde{C}=0,\ \tilde{y} > 0,
\eqno{(\theequation{\textrm{\textit{a,b}}})}\label{eq:purediffIC}
$$
and conditions of continuity at the drop--film interface for $\tilde{t} > 0$ satisfying
\refstepcounter{equation}
$$
\lim\limits_{\tilde{y}\rightarrow 0,\: \tilde{y}\leq 0} \tilde{C} = \lim\limits_{\tilde{y}\rightarrow 0,\: \tilde{y}\geq 0} \tilde{C}, \quad \textrm{and} \quad \lim\limits_{\tilde{y}\rightarrow 0,\: \tilde{y}\leq 0} D_d\frac{\p \tilde{C}}{\p \tilde{y}} = \lim\limits_{\tilde{y}\rightarrow 0,\: \tilde{y}\geq 0} D_f \frac{\p \tilde{C}}{\p \tilde{y}}.
\eqno{(\theequation{\textrm{\textit{a,b}}})}\label{eq:purediffBC}
$$
Note that the time and spatial coordinates have been left in their dimensional forms in  (\ref{eq:purediff}) to emphasize the change of diffusion coefficient between the two phases.  {The mass diffusion problem described above is analogous to the conduction of heat in an infinite one-dimensional composite solid \mbox{\cite[\eg][]{carslaw86}}}. According to \cite{carslaw86}, the solution to the problem (\ref{eq:purediff}) under the initial conditions (\ref{eq:purediffIC}) and the boundary conditions (\ref{eq:purediffBC}) is, for $\tilde{t} > 0$,
\begin{eqnarray}
\tilde{C} &  = & \frac{1}{1+\sqrt{\xi}}\(1+\sqrt{\xi} \erf\[\frac{\lvert h_d \tilde{y} \rvert}{2\sqrt{D_d t}}\] \),\ \tilde{y}\leq 0, \label{eq:purediffSolD} \\
 \tilde{C} &  = & \frac{1}{1+\sqrt{\xi}} \erfc\[\frac{h_d \tilde{y}}{2\sqrt{D_f t}} \],\ \tilde{y} \geq 0, \label{eq:purediffSolF}
\end{eqnarray}
where $\erf[\cdot]$ and $\erfc[\cdot]$ are the error function and the complementary error function, respectively. 
The solution (\ref{eq:purediffSolD}--\ref{eq:purediffSolF}) above is valid for $\tilde{t} < \tilde{t}_{dif}$. As $\tilde{t} \geq \tilde{t}_{dif}$, the establishment of the boundary layer in the film imposes the flux condition (\ref{eq:massfluxj}) at the drop--film interface. Therefore, as $\tilde{t}$ approaches $\tilde{t}_{dif}$ there should be a transition between the purely diffusive regime (\ref{eq:purediffSolD}--\ref{eq:purediffSolF}) with time-dependent transport in both the drop and the film, and the regime described by (\ref{eq:1ddiffNDsolavg}) with time-dependent transport in the drop and steady transport in the film.

We plot the concentration fields (\ref{eq:purediffSolD}) for $\tilde{t}=0.01$ and (\ref{eq:1ddiffNDsolsimp}) for $\tilde{t}\geq 0.1$  (plotted using different line styles), and the time evolution of the spatially averaged concentration (\ref{eq:1ddiffNDsolavg}) in figures~\ref{fig:DropCmodel}(\textit{a,b}), respectively. We use typical values for the different parameters: $\Rey_L=30\thinspace 000$, $\Sc_f=2000$, $\xi=5/3$, $\ell=\sqrt{\pi} L/2$ and $\eta_d=h_d/L=0.15$.

 {As we can see in \mbox{figure~\ref{fig:DropCmodel}(\textit{a})}, the concentration $\tilde{C}$ at the drop interface,  $\tilde{y}=0$, decreases very rapidly at early times, $\tilde{t} < 0.1$,  because diffusion is predominant and the diffusive boundary layer in the film is not fully developed. We note that the jump from $\tilde{C}(\tilde{y}=0,\tilde{t} =0.01)=1/2$ (see thin solid curve computed using \mbox{(\ref{eq:purediffSolD})}) to} $\tilde{C}(\tilde{y}=0,\tilde{t} =0.1)\approx 0.02$  {(see long-dashed curve computed using \mbox{(\ref{eq:1ddiffNDsolsimp})}) is partly accentuated by \mbox{(\ref{eq:purediffSolD})}, predicting $\tilde{C}(\tilde{y}=0)=1/2$ at all time due to the symmetry of the pure diffusive model \mbox{(\ref{eq:purediff})}. Thus, as advection processes are becoming more  important in the film phase during the transition described above at $\tilde{t} \approx 0.01$, \mbox{(\ref{eq:purediffSolD})} slightly overestimates the concentration at the interface. Then, as $\tilde{t} \approx 0.1$ the effect of the solid boundary at $\tilde{y}=-1$ becomes important, as shown by the long-dashed curve in \mbox{figure~\ref{fig:DropCmodel}(\textit{a})}.}

As shown in figure~\ref{fig:DropCmodel}(\textit{b}), the spatially averaged concentration decrease is nearly exponential in time. We  also plot in figure~\ref{fig:DropCmodel}(\textit{b}) a simple exponential decrease $\exp[-\pi^2\tilde{t}/4]$ with a dashed line. We can see that for $\tilde{t} \gtrapprox 0.1$,  $\tilde{C}_d$ decreases at the same rate as the simple exponential decrease. This shows that
for $\tilde{t}$ sufficiently large, $\tilde{t} \gtrapprox 0.1$, $\tilde{C}_d$ is dominated by the first term of the series $n=0$ in (\ref{eq:1ddiffNDsolavg}), with characteristic time 
\begin{equation}\label{eq:tauddiff2}
	\tau_{d,dif} \approx \frac{4 \mbox{$h_d$}^2}{\pi^2 D_d},
\end{equation}
which is equivalent to the drop diffusion time scale estimated previously in (\ref{eq:taudDif}). The  coefficient of proportionality in (\ref{eq:tauddiff2}) differs with (\ref{eq:taudDif}) due to the first parameter of the series $\lambda_0=\pi/2$ in (\ref{eq:1ddiffNDsolavg}).
This result strongly supports our Newton-cooling model which predicts that the mean concentration inside the drop decreases exponentially in time such that $\hat{C}_d=\exp[-t/\tau_{\kappa}]$ (see (\ref{eq:Csolution})).
By correspondence of the characteristic time scales, we must have 
\begin{equation}\label{eq:taukappaeqtauddif}
\tau_{\kappa}=\tau_{d,dif},
\end{equation}
with $\tau_{d,dif}$ described by (\ref{eq:tauddiff2}). This finding is valid at least for low drop \Pns, $Pe_d\ll 1$, and corresponds to the diffusion-dominated regime (D) presented in (\ref{eq:regimeD}). In regime D, the overall time scale for the mass transfer across the drop--film interface is influenced mainly by the diffusive transport inside the drop. In this regime, the film phase has a limited influence on the change of the concentration inside the drop through either the  Reynolds number $\Rey_L=\gamma L^2/\nu_f$, the Schmidt number $\nu_f/D_f$ or the diffusivity of the tracer in the film phase $D_f$.

For drop \Pns of order $1$, which corresponds to the drop advection--diffusion regime (AD) presented in (\ref{eq:regimeAD}), the transport inside the drop is governed by a time-dependent advection--diffusion equation.  {Diffusive transport occurs throughout the drop for  $\Pe_d\approx 1$ in combination with an advective recirculation process. As $\Pe_d$ increases, the effect of diffusion becomes limited to a thin diffusive boundary layer below the drop--film interface. In regime AD,} the characteristic transport time  must be of the same order of both the advection and diffusion time scales, by definition,
\begin{equation}\label{eq:taud_taudif_tauadv}
	\tau_{d} \approx \tau_{d,dif} \approx \tau_{d,adv} \approx  10^2 \; \textrm{s}.
\end{equation}
While the full time-dependent advection--diffusion equation in the drop is not solved in the present study, we discuss further the implications of advection processes inside the drop on the overall mass transfer when showing the experimental results in section \ref{sec:overallmasstransfer}.

\subsection{Sherwood number}\label{sec:sherwoodnum}

We can now fully determine theoretically the convective mass transfer coefficient $\kappa$. From the Newton-cooling model, we can define a Sherwood number, or non-dimensional convective mass transfer coefficient, such that
\begin{equation}\label{eq:defShkappa}
\overline{Sh} = \frac{\kappa \ell}{D_f},
\end{equation}
with  {$\ell=\sqrt{A}$, and} $\kappa$ the spatially averaged normalised flux defined as
\begin{equation}\label{eq:defkappa}
\kappa = \frac{\lvert F \rvert}{A(C_d-C_{\infty})},
\end{equation}
where $\lvert F \rvert$ is the magnitude of the total mass flux through the drop--film interface of area $A$, $C_d$ is the spatially averaged concentration in the drop, and $C_{\infty}$ is the background concentration outside the drop.
By inspection of $\overline{Sh_f}$ in (\ref{eq:defSh}) and since $F=F_f$, we have
\begin{equation}\label{eq:ShShf}
\overline{Sh} = \frac{(C_{f,i}-C_{\infty})}{(C_d-C_{\infty})} \overline{Sh_f},
\end{equation}
where $C_{f,i}$ and $C_d$ can be determined independently from the model of the mass transport in the drop phase using (\ref{eq:1ddiffNDsolsimp}) and (\ref{eq:1ddiffNDsolavg}), respectively.  {In the drop diffusive regime (D),} we find that the ratio $(C_{f,i}-C_{\infty})/(C_d-C_{\infty})$ is constant for $\tilde{t}$ sufficiently large. Therefore, for $\tilde{t}>0.1$, (\ref{eq:ShShf}) becomes
\begin{equation}\label{eq:ShvsRe13}
\overline{Sh} = \overline{\Sh}_{model,D} \approx \frac{0.766 \mbox{$\Rey_L$}^{1/3} \mbox{$\Sc_f$}^{1/3}}{1+ 0.766 \mbox{$\Rey_L$}^{1/3} \mbox{$\Sc_f$}^{1/3} \frac{8}{\pi^{5/2} } \eta_d  \xi},
\end{equation}
with the characteristic Reynolds number $\Rey_L=\gamma L^2/\nu_f$, the Schmidt number in the film $ \Sc_f=\nu_f/D_f$, the drop aspect ratio $\eta_d = h_d/L$, and the diffusivity ratio $\xi=D_f/D_d$.  {We stress that the theoretical prediction \mbox{(\ref{eq:ShvsRe13})} is a priori valid only in regime D for $Pe_d \ll 1$.}

It is interesting to note that for large Reynolds numbers, Schmidt numbers or diffusivity ratio, the Sherwood number can reach an asymptotic limit:
\begin{equation}\label{eq:Shmax}
\overline{Sh} \rightarrow \overline{Sh}_{max} = \frac{\pi^{5/2}}{8\eta_d \xi} = \frac{\pi^{5/2}L D_d}{8 h_d D_f} \ \ \textrm{for} \ \(\mbox{$\Rey_L$}^{1/3} \mbox{$\Sc_f$}^{1/3}  \eta_d \xi \) \gg 1. 
\end{equation}
Therefore, in regime D the convective mass transfer coefficient has a theoretical maximum $\overline{Sh}_{max}$, which only depends on the drop aspect ratio $\eta_d$ and the diffusivity ratio $\xi$.  {The result \mbox{(\ref{eq:Shmax})} assumes that the volume flux in the film is not important.}

On the other hand, we can see that if $(\mbox{$\Rey_L$}^{1/3} \mbox{$\Sc_f$}^{1/3}  \eta_d \xi )\ll 1$, then the Sherwood number corresponds to the more familiar regime 
\begin{equation}\label{eq:Shwellmixed}
\overline{Sh} = \overline{Sh_f} = 0.766 \mbox{$\Rey_L$}^{1/3} \mbox{$\Sc_f$}^{1/3}. 
\end{equation}
This regime was identified in previous studies for the case of a saturated drop containing only one species \cite[][]{stone89,baines:james94,danberg08,blount10}. This regime effectively corresponds to a well-mixed drop, which can occur if tranport inside the drop is much faster than the transport from the interface to the bulk of the film, \ie $\tau_d \ll \tau_f$. Indeed, for a well-mixed uniform drop we have $C_{f,i}=C_d$, and according to (\ref{eq:ShShf}) $\overline{Sh} = \overline{Sh_f}$, hence (\ref{eq:Shwellmixed}).  {The} model presented in section \ref{sec:modeldropphase} was formulated for the case where diffusion dominates transport in the drop. Nevertheless, this model does predict that for $(\mbox{$\Rey_L$}^{1/3} \mbox{$\Sc_f$}^{1/3}  \eta_d \xi )\ll 1$ the drop can appear well mixed and the Sherwood number reaches the saturated or well-mixed regime described by (\ref{eq:Shwellmixed}).

In the transition from a non-uniform diffusive drop to a well-mixed drop, the Sherwood number probably evolves  from equation (\ref{eq:ShvsRe13}) to (\ref{eq:Shwellmixed}). We surmise that this transition occurs smoothly as the diffusive boundary layer forming below the drop--film interface, in the drop phase, becomes thinner and the concentration jump $(C_{f,i}-C_{\infty})/(C_d-C_{\infty})$ decreases. According to (\ref{eq:ShShf}), if the concentration jump between the drop bulk concentration $C_d$ and the interfacial concentration $C_{f,i}$ reduces, then the Sherwood number approaches the well-mixed regime $\overline{Sh} = \overline{Sh_f}$. It is possible that this transition occurs in regime AD ($\Pe_d\approx 1$), as advection processes start reducing the thickness of the diffusive boundary layer in the drop phase.

We can compare our results for the Sherwood number  {with past results from the  literature. As mentioned in the introduction, all past results} are for well-mixed drops, or for cases in which only the concentration at the interface is considered and assumed constant. \cite{stone89} found that the Nusselt number varies such that $\Nu = 2.157\Pe^{1/3}+3.55\Pe^{-1/6}$  {for the heat transfer from a flat disk on a plane boundary in a uniform shear flow.} Thus, at high  \Pns, the result of \cite{stone89} is equivalent to our relationship (\ref{eq:Shwellmixed}) in the well-mixed regime. 
\cite{baines:james94} found the same mass flux as in (\ref{eq:massfluxj}). Assuming $h_d=0$, they obtained $\overline{Sh} = 0.745 \mbox{$\Rey_L$}^{1/3} \mbox{$\Sc_f$}^{1/3}$, using our definition (\ref{eq:defShkappa}) for the Sherwood number. The coefficient of proportionality differs slightly with  (\ref{eq:Shwellmixed}) because \cite{baines:james94} reshaped the surface area of the drop into a square. \cite{danberg08} also obtained a very similar result: $\overline{Sh} = 0.755 \mbox{$\Rey_L$}^{1/3} \mbox{$\Sc_f$}^{1/3}$, using our definition (\ref{eq:defShkappa}). This result also differs slightly with (\ref{eq:Shwellmixed}) because \cite{danberg08} assumed a third-order polynomial distribution for the concentration field in the boundary layer instead of solving it from the advection--diffusion equation. We summarize past and current results for the spatially averaged Sherwood number in table~\ref{tab:ShAll}, using both  original and  current notations.

\begin{table}
	\begin{minipage}{\textwidth}
		\centering
		\begin{tabular}{@{}lcc@{}}
			\hline \\
			{Source} & {Original notations} & {Current notations}  \\[3pt]
			\cite{stone89} & $\Nu = 2.157\Pe^{1/3}+3.55\Pe^{-1/6}$ &  --  \\
			\cite{baines:james94} & $Sh = 0.105 \mbox{$\Pe$}^{1/3}$ &  $\overline{Sh} = 0.745 \mbox{$\Rey_L$}^{1/3} \mbox{$\Sc_f$}^{1/3}$  \\
			\cite{danberg08} &  $\overline{Sh_L} = 0.852 (u_{\tau} L/\nu)^{2/3} \mbox{$\Sc$}^{1/3}$ & $\overline{Sh} = 0.755 \mbox{$\Rey_L$}^{1/3} \mbox{$\Sc_f$}^{1/3}$ \\ \midrule[0.01pt]
			\multicolumn{2}{l}{Present study (well-mixed drop)} &  $\overline{Sh} = \(0.766+1.131 \mbox{$\eta_d$}^2 \) \mbox{$\Rey_L$}^{1/3} \mbox{$\Sc_f$}^{1/3}$  \\
			\multicolumn{2}{l}{Present study (non uniform drop with $\Pe_d\ll 1$)} &  $\overline{Sh} = \frac{\displaystyle 0.766 \mbox{$\Rey_L$}^{1/3} \mbox{$\Sc_f$}^{1/3}}{{\displaystyle 1+ 0.766 \mbox{$\Rey_L$}^{1/3} \mbox{$\Sc_f$}^{1/3}} { \frac{8}{\pi^{5/2} }} {\displaystyle \eta_d  \xi}}$  \\	\\ \hline
		\end{tabular}
	\end{minipage}
	\caption{Results of past and current studies for the Sherwood number (or Nusselt number) using the original notations, and computed using our current notations with the definition (\ref{eq:defShkappa}) for the Sherwood number. The mass transfer occurs from a thin drop lying on a flat substrate and surrounded by an external shear flow (see figure~\ref{fig:droplet-film}). The  Reynolds number in the shear flow is: $\Rey_L=\gamma L^2/\nu_f$. The friction velocity $u_{\tau}$ used by \cite{baines:james94} is equivalent to $\sqrt{\nu_f \gamma}$ for our study.}
	\label{tab:ShAll}
\end{table}

\section{Experimental procedure}\label{sec:xpproc}

We conducted experiments to measure the concentration of methylene blue dye used as tracer in a polymer-thickened droplet submerged in a thin falling film of water ( {see} figure~\ref{fig:droplet-film}).  {The film Reynolds number at this location ranged: $\Rey_f=h_{\infty} U_f/\nu_f=500$ to $1100$, with $h_{\infty}$ the thickness of the fully developed unperturbed film.} Undisturbed, we assumed that the  far-field velocity profile of the film was well approximated by a viscous--gravity balance (see equation (\ref{eq:viscousgravU})), even in this intermediate range of Reynolds number \cite[see][]{landel15}. 

The droplets contained a non-ionic water-soluble polymer Natrosol 250 hydroxyethylcellulose (HEC) produced by Aqualon. HEC is a non-Newtonian shear thinning polymer. We used the most viscous  {type:} 250HHR (molecular weight: $1.3 \times 10^{6}$; density: 1.0033; $\textrm{pH} = 7$, refractive index: $1.336$) at $2$\%~wt concentration in tap water.  {According to our rheological measurements (see \mbox{Appendix~\ref{apx:RheoDrop}}), the drop phase was always in a Newtonian regime with characteristic viscosity $\mu_d\approx 100\;$Pa s. For film shear rates $\gamma\approx 10^3\;$s$^{-1}$, the typical shear rate in the drop phase was approximately $10^{-2}\;$s$^{-1}$. We chose a very viscous polymer in order to reduce the  amplitude $\varphi=A_m/A_0$ of deformation of the droplets induced by the film shear (see \mbox{Appendix~\ref{apx:Aint})} and separate the characteristic time scales such that $\tau_{\delta}\ll \tau_{d,def} \ll \tau_{\kappa}$ in most experiments (see table~\mbox{\ref{tab:expcharactime}}).} Without methylene blue dye, the polymer solution was colourless and almost as clear as  {water.} 

 {We measured the diffusivity of the methylene blue dye (Fisher Scientific) in both the drop phase (aqueous $2$\%~wt polymer solution) and the film phase (tap water). Conducting these experiments in capillary tubes (see \mbox{Appendix~\ref{apx:DiffExp}}) at $20^{\circ}$C, we found in the drop and film phases, respectively, $D_d=3 \times 10^{-10}\:$m$^2\:$s$^{-1}$ and $D_f=5 \times 10^{-10}\:$m$^2\:$s$^{-1}$. Our measurements are consistent with the measurements of \mbox{\cite{sedlacek13}} who found $D=8.4 \times 10^{-10}\:$m$^2\:$s$^{-1}$ in pure water at $25^{\circ}$C and $D=2.9 \times 10^{-10}\:$m$^2\:$s$^{-1}$ in hydrogel solutions  at $25^{\circ}$C. The smaller diffusivity in the drop phase (constituted mainly of water) compared with the pure water film phase could be due to electro-chemical interactions between the methylene blue dye ions and some parts of the HEC polymer chain.}

\begin{figure}
\centering
\begin{tikzpicture}[scale=0.55]
\draw[semithick] (0,5) -- (5,0)  ++(45:0.2) -- ++(135:7.07);
\draw (5,0) ++(135:2) arc (135:180:2);
\draw (5,0) -- (1.3,0);
\draw (2.85,0.85) node {$\alpha$};
\draw[semithick] (0,5) ++(-45:3) ++(45:0.4) -- ++(-45:0.2) -- ++(45:1.5) -- ++(135:0.2) -- ++(-135:1.5);
\draw[semithick] (0,5) ++(-45:2.5) ++(45:0.5) -- ++(45:0.2) -- ++(135:0.6) -- ++(90:0.8);
\draw[thin,pattern=dots] (0,5) ++(-45:2.45) ++(45:0.25) -- ++(45:0.4) -- ++(135:0.7) -- ++(-135:0.4) -- cycle;
\draw[semithick] (0.9,4.7) -- ++(-65:0.5); \draw[semithick] (0.9,4.7) ++(25:0.15) -- ++(-65:0.5);
\draw[<-] (0.9,4.7) ++(45:0.075) -- ++(115:0.3) node[above] {\footnotesize Pump};
\draw[dashed] (0,5) ++(-45:1.6) ++(45:0.2) -- ++(0:2);
\draw[->] (0,5) ++(45:0.05) ++(-45:3.5) -- ++(45:0.15);
\draw[->] (0,5) ++(45:0.55) ++(-45:3.5) node[right] {\scriptsize $h_0$} -- ++(-135:0.15);
\draw (0,5) ++(45:0.4) ++(-45:3.2) -- ++(-45:3.5) arc (45:-45:0.2);
\draw (0,5) ++(45:0.2) ++(-45:4.5) arc (135:-45:0.1);
\draw[->] (1,3.7) node[left] {\footnotesize Foam} -- ++(25:0.45);
\draw[->] (1,3) node[left] {\footnotesize Solid substrate} -- ++(35:0.7);
\draw[->] (3.7,5.5) node[right] {\footnotesize Distribution manifold} -- ++(-145:2.25);
\draw[->] (5,4) node[right] {\footnotesize Reservoir} -- ++(-175:1.65);
\draw[->] (5,2.5) node[right] {\footnotesize Viscous droplet} -- ++(-160:1.6);
\draw[->] (5,1.5) node[right] {\footnotesize Film} -- ++(-160:0.7);
\draw (6,0.5) node[right] {\footnotesize Study area};
\end{tikzpicture}
~
\begin{tikzpicture}[scale=0.65]
\draw[semithick] (0,0) rectangle +(4,5);
\draw[semithick] (0.5,4) -- (0.5,2) -- (3.5,2) -- (3.5,4) -- (3.3,4) -- (3.3,2.2) -- (0.7,2.2) -- (0.7,4) -- cycle;
\draw (0.7,3.5) -- (3.3,3.5);
\draw[dashed] (1,2.2) -- (1,2); \draw (1,2) -- ++(-90:1.7) arc (-180:-90:0.2) -- ++(0:1.6) arc (-90:0:0.2) -- ++(90:1.7); \draw[dashed] (3,2.2) -- (3,2);
\draw (2,1.1) arc (0:360:0.04);
\draw[->] (-0.7,1.9) -- (1.92,1.12);
\draw[thin, dashed] (1.4,0.6) rectangle +(1.2,1.2);
\draw[->] (-0.5,0.5) -- (1.4,0.6);
\end{tikzpicture}
\caption{Schematic  of the experimental apparatus: (\textit{a}) side view; (\textit{b}) top view. The camera records the flow in the study area. This is the same apparatus as used by \cite{landel15}.}
\label{fig:Exp-app-fig}
\end{figure}
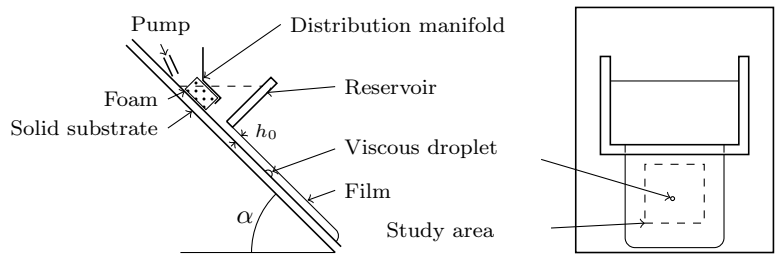

 {We  conducted the main experiments using the apparatus shown schematically in \mbox{figure~\ref{fig:Exp-app-fig}}; this is the same  as used by \mbox{\cite{landel15}}. A liquid film of tap water flowed from a constant-head reservoir through a thin gap $200\:$mm wide and $15\:$mm long in the streamwise direction. The gap thickness $h_0$ could be adjusted using a micrometre screw. At the exit of the gap, the film flowed on a flat glass substrate inclined at an angle $\alpha$ from the horizontal. We measured the angle of inclination $\alpha$ using an electronic inclinometer (precision better than $0.1^{\circ}$). 
The $10\:$mm thick glass plane used as substrate ensured the rigidity of the experimental apparatus. The  substrate was cleaned before each experiment with water and soap, then isopropanol, to maintain consistent wetting properties. The film flowed freely on the substrate for a distance of approximately $300\:$mm from the outlet of the reservoir gap to the bottom-end of the substrate, and then fell  into a large collecting tank.
The fluid  recirculated in the  apparatus using a  pump located in the collecting tank.} The fluid was pumped into a distribution manifold located upstream of the main reservoir. The fluid turbulence in the distribution manifold was dampened as it penetrated through a piece of foam (a reticulated polyether foam with $57$ to $70$ pores per inch and a pore size of approximately $0.5\:$mm) and a $5\:$mm high two-dimensional funnel into the main U-shaped reservoir (see figure~\ref{fig:Exp-app-fig}). Using some artificial pearlescence (Iriodin 120 pigment, Merck) we observed that the fluid in the main U-shaped reservoir was free of turbulence.
The total flow rate of the film was obtained by measuring the flow rate falling  from the end of the substrate. Using a precision balance (to measure the mass of fluid) and a stopwatch for each experiment, the flow rate was found to be consistent through repeated measurements with a precision of approximately  $1$\%.

\begin{figure}
	\centering
	\includegraphics[trim = 65mm 40mm 70mm 40mm, clip, width=0.45\textwidth]{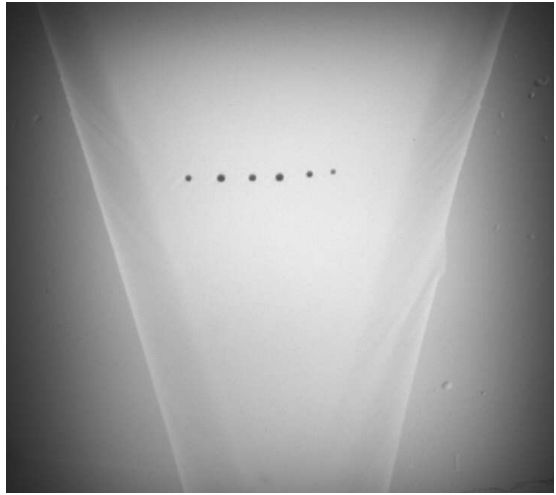}
	\caption{Camera view from above at right angle with the substrate. Six droplets of $2$\%~wt HHR polymer and methylene blue dye are identified as small black dots in the interior of the film. The film contraction due to surface tension is visible at the edges. The ropes, or thicker edges, of the film can be seen in slightly darker grey than the interior.}
	\label{fig:drops-film-pic}
\end{figure}

As found by \cite{landel15}, the velocity in the film flow surrounding the droplets was essentially downstream with a very small cross-stream component (of the order of $5$\% of the far-field streamwise velocity component). The streamwise surface velocity could decrease by $20$ to $50$\% as the film flowed over the droplet, depending on the ratio between the film height and the droplet height. We also observed stationary capillary waves forming on the sides of the droplets in a bow-wave pattern  \mbox{\cite[][]{gaskell04}}.  {The contraction of the film flow, due to surface tension and which can be observed in  \mbox{figure~\ref{fig:drops-film-pic}}, was mostly accommodated by an increase in volume flux in the ropes seen at the edges of the film \mbox{\cite[][]{landel15}}.  The ropes are the faster and thicker edges of the flow \mbox{\cite[][]{lan10}}.}

Several drops containing the mixture of polymer solution and methylene blue dye were aligned across the substrate approximately $8$ to $10\:$cm downstream from the source reservoir where the film reached its fully-developed state \cite[][]{landel15}.  {The typical drop had length $L\approx 1$ to $4\;$mm and height $h_d\approx 0.3$ to $0.6\;$mm. To minimize the effect of evaporation of  the droplets, we covered them with the water film within $30\:$s after deposition}. This  time  also allowed  the droplets to reach a sessile shape and increase their adherence to the substrate. 

To ensure a rapid and simultaneous cover of all the droplets by the film we used the following protocol, which also limited the impact of the fingering instability associated with the advancing contact line \cite[][]{huppert82}. Before starting the flow, we protected the region where the droplets lay by placing a barrier to divert the flow around both sides of the drops. Once the film wetted the surrounding area and reached its fully-developed regime, we removed the barrier to let the film submerge the droplets. This protocol guaranteed a precise reference for the starting time of an experiment $t=0$, corresponding to the moment when the film covered  the droplets.

We show in figure~\ref{fig:drops-film-pic} a photograph of a thin film flowing over six drops, seen as small black dots  {lying  in} the interior of the film. The distance between each drop was approximately $3$ to $4$ drop diameters, which was sufficient to prevent any interference in the flow over and around the neighbouring drops \cite[][]{landel15}. Moreover, we  laid the drops in the uniform flow region in the interior of the film, at least $2\;$cm away from  {the ropes. This distance was deemed sufficient based on our analysis of the height profile across the film width at the location of the drops.} We  noticed that stationary surface capillary waves could appear at the surface of the film at the highest Reynolds numbers, propagating downstream at an angle of approximately $55^{\circ}$ from the horizontal. These capillary waves were produced by very small surface defects (such as small dust on the surface or changes in surface chemistry) disturbing the side contact line of the film. The amplitude of these capillary waves was very small in the interior of the film: of the order of a few tens of microns. The film  did not show any of the well-known long wave instabilities in the regimes studied \cite[\eg][]{kalliadasis12}.

 {We used the dye attenuation technique to measure the time evolution of the concentration of methylene blue dye, $C_{MB}$, in the polymer droplets submerged in the film. We first calibrated the 8-bit greyscale camera (Jai CVM4+CL, mounted with a $75\;$mm lens, aperture: f/8.0D) to ensure that it had a linear response with light intensity and to measure the value of its black offset. We used an array of $6 \times 9$ red LEDs (TruOpto) in combination with a diffusive white acrylic sheet to produce a constant, uniform light source. This light source was located approximately $20\:$cm behind the back of the glass substrate. The peak wavelength of the LEDs (luminous intensity: $0.5\:$cd each), $625\:$nm, was close to the absorption peak wavelength of methylene blue: $664\;$nm. We measured the transmitted light intensity with the camera described above, located between $0.5\;$m and $0.8\;$m  perpendicularly away from the substrate, which was sufficient to have negligible parallax error. We performed the experiments in a dark room and took care to reduce any light pollution from reflection or other sources. The camera recorded $1360 \times 1030$ pixel images. The acquisition rate was set at $24$ frames per second for a duration of  $3$ to $8$ minutes, until no dye could be seen by the detection system.}

 {To prepare the dye attenuation technique, we followed the calibration method and the algorithm described by \mbox{\cite{allgayer12}} and based on \mbox{\cite{cenedese98}}. We performed  the calibration in situ. We fixed the dye concentration in the liquid recirculating in the  apparatus and captured images of the dyed  film. The camera recorded the intensity for a certain local film thickness. We measured the film thickness in the undisturbed far-field flow (\mbox{\ie} without droplets) using a high-precision digital micrometre (Mitutoyo, accuracy of $10^{-6}\;$m, $1\:$mm blunt tip) at the location of the droplets.
The height of the film was  measured at a precision of approximately $10^{-5}\;$m. The images recorded by the camera were analysed using the software code DigiFlow \mbox{\cite[][]{dalziel07}}. We obtained an accurate relationship between the intensity recorded by the camera and the depth-averaged concentration in the film.
Even though the calibration was performed for thin flat films, it was also valid for the measurement of the methylene blue dye in a polymer droplet. Once the droplets were submerged, owing to the similar refractive indices for the polymer solution (1.336) and water (1.333), the effect of the droplet curved surface on the path of the light rays for the dye attenuation was negligible.}

 {We conducted $5$ different sets of experiments, for a total number of $119$ drops.  All the experiments were conducted at room temperature, $20^{\circ}$C, and the solutions were at $\textrm{pH} = 7$. We present in \mbox{table~\ref{tab:expcondfilm}} the experimental conditions related to the film flow: $\alpha$ is the substrate inclination angle, $h_0$  is the gap thickness of the reservoir, $h_{\infty}$ is the undisturbed film thickness at the location of the droplets as measured with the high-precision micrometre, $Q$ is the volume flow rate, $U_f$ is the theoretical depth-averaged film velocity, and $\Rey_f=h_{\infty}U_f/\nu_f$  is the fully-developed film Reynolds number. }

In half of the experiments, some ascorbic acid was added to the liquid in the film in order to study the impact of reaction processes on the mass transfer. Ascorbic acid can react with methylene blue to form a colourless compound \cite[][]{mowry99}. At the concentration used for the results reported here, the experiments with ascorbic acid did not show any significant difference with the other experiments. We choose to include these experiments to extend the parameter regime covered in the present study.

\begin{table}
  \begin{minipage}{\textwidth}
    \centering
    \begin{tabular}{@{}ccccccc@{}}
    	\hline \\
      {Exp.} & {Angle $\alpha$ ($^{\circ}$)} & {$h_0$ (mm)} & {$h_{\infty}$ (mm)} & {$Q$ (\SI{}{cm^3. s^{-1}})} & {$U_f$ (\SI{}{m. s^{-1}})} & {$\Rey_f$} \\[3pt]
      $1$ & $20.0$ & $0.40$ & $0.8$ & $47$ & $0.7$ & $500$ \\
      $2$ & $30.0$ & $0.40$ & $0.7$ & $47$ & $0.8$ & $600$ \\
      $3$ & $30.0$ & $1.00$ & $0.9$ & $91$ & $1.3$ & $1100$ \\
      $4$ & $45.0$ & $0.40$ & $0.6$ & $46$ & $1.0$ & $600$ \\
      $5$ & $45.0$ & $0.60$ & $0.7$ & $64$ & $1.1$ & $800$ \\ \\ \hline
    \end{tabular}
  \end{minipage}
\caption{Experimental conditions related to the film flow.}
\label{tab:expcondfilm}
\end{table}

 {In \mbox{table~\ref{tab:expconddrop}}, we present the experimental parameters related to the droplets. All the experiments were conducted with an initial methylene blue dye concentration of} $C_0 = \SI{0.1}{kg. m^{-3}}$ ($0.01$\%~wt)  {so that the response of the camera to variations of concentration in the droplets was approximately linear. Owing to some noise in the experiments and the 8-bit quantisation of the images, the lowest concentration measured by our experimental setup was $C_{det}/C_0 = 0.05$ to $0.09$. The smallest area detected by the camera was  $A_{det}$, which was of the order of $5$ pixels or approximately $0.2\:$mm$^2$. Initial drop volumes before the film covered the drops were from $0.5$ to $5\:$mm$^3$ and initial drop areas were from $2$ to $7\:$mm$^2$, as measured by our dye detection system.}

\begin{table}
  \begin{minipage}{\textwidth}
    \centering
    \begin{tabular}{@{}ccccc@{}}
    	\hline \\
      {Exp.} & {Number of drops} & {$C_0$ (\%~wt)} & {$C_{det}/C_0$} & {$A_{det}$ (mm$^2$)}\\[3pt]
      $1$ & $22$ & $0.01$ & $0.06$ & $0.185$ \\
      $2$ & $30$ & $0.01$ & $0.06$ & $0.274$ \\
      $3$ & $25$ & $0.01$ & $0.06$ & $0.274$ \\
      $4$ & $21$ & $0.01$ & $0.05$ & $0.175$ \\
      $5$ & $21$ & $0.01$ & $0.09$ & $0.175$ \\ \\ \hline
    \end{tabular}
  \end{minipage}
\caption{Experimental conditions related to all the drops studied.}
\label{tab:expconddrop}
\end{table}

\section{Experimental results}\label{sec:xpresults}

\subsection{Phenomenological description of drops submerged in thin falling film}

\begin{figure}
	\centering
	\vspace{0.3cm}
	\def\svgwidth{1\textwidth}
	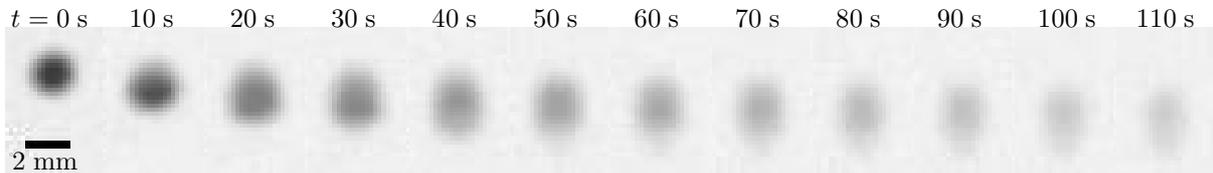
	\caption{ {Time sequence of greyscale images (view from above) of a polymer drop mixed with methylene blue dye, as seen by the camera as a water film flows over the drop for $t\geq 0$.}
	}
	\label{fig:Alldrops}
\end{figure}

We show in figure~\ref{fig:Alldrops} a time sequence ($0\leq t \leq 110\;$s) of greyscale images of a typical polymer drop with methylene blue dye (fading from black to grey) submerged in a water film. These images are close-up views taken from Exp. 4 ($\Rey_f=600$). The film submerges the drop at $t=0$. The centre of mass of the drop displaces downstream (\ie downwards) in the first $20$ to $30\:$s. The drop deforms and elongates in the streamwise direction due to the shear imposed by the falling film flow. As the drop elongates, it also becomes thinner. The drop appears to reach a fixed position at $t\approx 30\:$s. 
We can also notice that the dyed area, as seen by the camera, increases in the first $30\:$s. We have measured increases in the area of up to $80$\% in our experiments. We have also computed the asymptotic drop aspect ratio $\eta_d$ from the initial drop volume $V_0$ and the asymptotic drop area $A_m$, and by assuming that the drop has a spherical cap shape. We find that $\eta_d$ ranges from  $0.11$ to $0.21$. These estimations validate our hypothesis that the drops are thin.

 {Throughout} an experiment the dye diffuses out of the drop: this can be seen in figure~\ref{fig:Alldrops} with the slow increase in the intensity of the drop, which changes from dark grey to light grey pixels. The dye detection system measures a depth-integrated value of the concentration using the dye attenuation technique described in the previous section. 

It is interesting to note that the intensity changes faster at the edge of the drop than at its centre. This is probably due to the fact that the thickness of the drop decreases noticeably at the edge. It could also mean that the mass transfer is faster at the  edge than at the centre, as discussed by \cite{stone89}, due to a discontinuity of the mass flux.

 {In all our experiments, the droplet volume remained approximately constant with $V(t) \approx V_0$. We measured qualitatively the diffusivity of the polymer solution in water and found that it was several orders of magnitude smaller than the diffusivity of the dye. Moreover, the shear in the film was not large enough to remove any drop material.}

Our aim is to test the Newton-cooling model presented in (\ref{eq:NewtonCooling}), as well as the advection--diffusion model coupled between the drop and film phases presented in sections \ref{sec:transfilm} and  {\mbox{\ref{sec:modeldropphase}}. 
For each experiment, we measured the amount of methylene blue dye} within a drop from the images shown in figure~\ref{fig:Alldrops}. To calculate the total mass of dye in the drop at each time instant (corresponding to a particular frame of the video), we computed the following formula
\begin{equation}\label{eq:massMBxp}
\mathcal{M}_{MB}(t) = \int\!\!\!\int_{A(t)} \(\int_{0}^h C_{MB}(x,y,z,t) \dd y\) \dd x \dd z,
\end{equation} 
where the depth-integrated concentration $\int_{0}^h C_{MB}(x,y,z,t) \dd y$ was obtained from the dye attenuation  {technique explained in \mbox{section~\ref{sec:xpproc}}. The area $A(t)$ is the dyed area seen by the detection system. Note $\mathcal{M}_{MB}(t)$  includes any dye present in the film above the drop,  hence the use of the film height $h$ instead of $h_d$ in the inner integral of \mbox{equation~(\ref{eq:massMBxp})}. This approximation has a negligible impact on $\mathcal{M}_{MB}(t)$. Indeed, we estimated in \mbox{section~\ref{sec:charatimescales}} that the diffusive boundary layer thickness above the drop was of the order of $\delta\approx 10^{-6}$ to $10^{-5}\;$m, which is much smaller than the drop thickness  $h_d\approx 0.3$ to $0.6\;$mm.}

\subsection{Overall mass transfer}\label{sec:overallmasstransfer}

\begin{figure}
	\centering
	\input{./VC_C0_no_corrLog_tnd-OverAll-JFM}
	\includegraphics[width=0.48\textwidth]{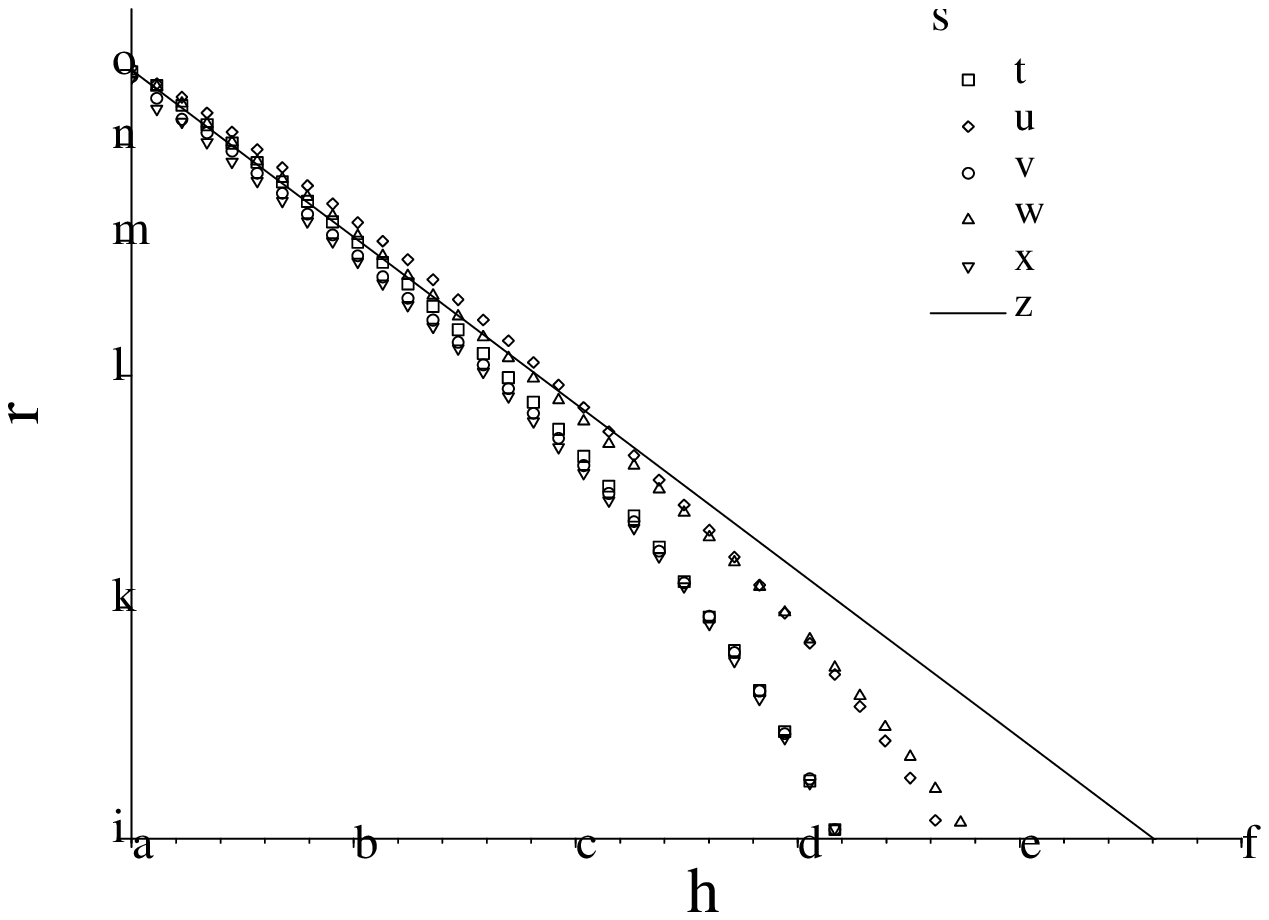}~
	\input{./VC_C0Log_tnd-OverAll-JFM}
	\includegraphics[width=0.48\textwidth]{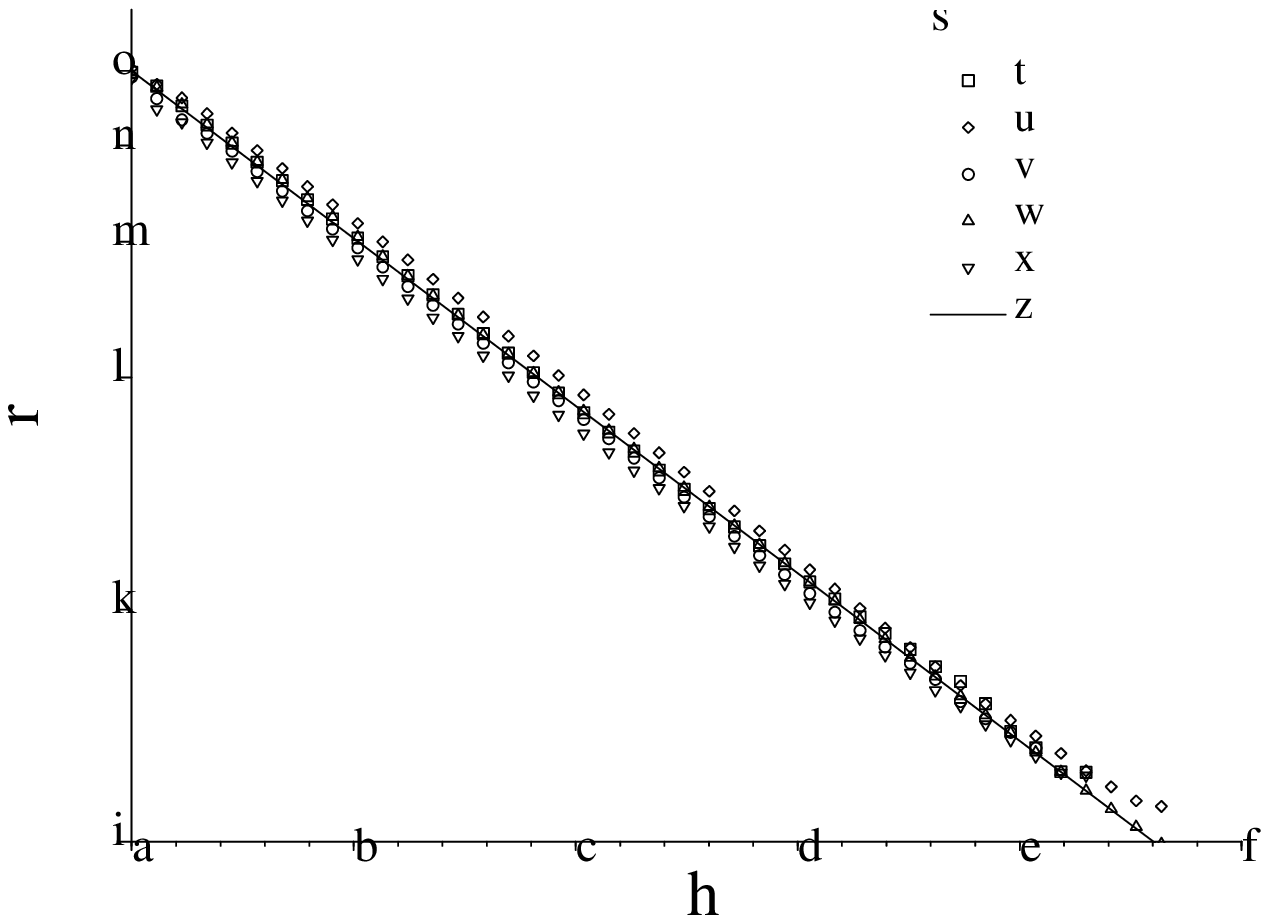}
	\caption{Evolution of the normalised quantity of dye $m_{MB}/m_{MB,0}$ on a logarithmic axis with non-dimensional time $t/\tau_{\kappa}$. The time scale $\tau_{\kappa}=V_0/(\kappa A_m)$ is the characteristic time for the convective mass transfer. We plot the ensemble-averaged data of the drops for each experiment using different symbols (see tables~\ref{tab:expcondfilm} and~\ref{tab:expconddrop}). The solid lines represent the exponential decrease predicted by the Newton-cooling model (\ref{eq:NewtonCooling}). In (\textit{a}) we show the data before  correction for the dye invisible to our detection system  {(see \mbox{(\ref{eq:massMBxp}))}, in (\textit{b}) we show the data after correction according to \mbox{(\ref{eq:corrdyedata})}}.
	}
	\label{fig:VC_C0_tnd-OverAll-JFM}
\end{figure}

In figure~\ref{fig:VC_C0_tnd-OverAll-JFM}(\textit{a}), we plot the ensemble average for each experiment of the normalised experimental data $\mathcal{M}_{MB}/\mathcal{M}_{MB,0}$  in time. Different symbols correspond to different experiments (see tables~\ref{tab:expcondfilm} and~\ref{tab:expconddrop}). The same symbols are used in figures~\ref{fig:VC_C0_tnd-OverAll-JFM} to~\ref{fig:ShL_ReL-OverAll-JFM} to designate the same experiments. The mass $\mathcal{M}_{MB,0}$,  {computed using \mbox{(\ref{eq:massMBxp})}, is the initial mass of methylene blue dye in the drop at $t=0$. If we assume that the concentration of methylene blue in the background is negligible, $C_{\infty}\approx 0$,} we note that $\mathcal{M}_{MB}/\mathcal{M}_{MB,0}$ is equivalent to the normalised drop concentration $\hat{C_d}$ in (\ref{eq:normCt}\textit{a}). 

The data in figure~\ref{fig:VC_C0_tnd-OverAll-JFM}(\textit{a}) are shown on a logarithmic axis for the normalised concentration $\mathcal{M}_{MB}/\mathcal{M}_{MB,0}$ against  non-dimensional time $\hat{t}=t/\tau_{\kappa}$, where $\tau_{\kappa}=V_0/(\kappa A_m)$ (see equations (\ref{eq:normCt}\textit{b}) and (\ref{eq:kappadef})). The parameter $\kappa$ is the convective mass transfer coefficient. We use $\kappa$ as fitting parameter to fit our experimental data for each of the $119$ drops with the Newton-cooling model $\mathcal{M}_{MB}/\mathcal{M}_{MB,0}=\exp[-t/\tau_{\kappa}]$. Using a least-squares fit, we find that $\kappa$ varies from $2.5\times 10^{-6}$ to $4.8\times 10^{-6}$\mpers. We analyse later the dependence of $\kappa$ with geometric and flow parameters, such as the Reynolds number $\Rey_L$. As explained in section~\ref{sec:masstransModel}, we use the maximum area $A_m$ reached by the droplet at $t\approx \tau_{d,def}$, instead of the time-dependent area $A(t)$. This simplification has a negligible effect on our results, because the characteristic time scale associated with the change of the area in time $\tau_{d,def}$ is much smaller than the characteristic time scale associated with the change of the dye concentration in the droplet $\tau_{\kappa}$ (see table~\ref{tab:expcharactime}, and  {\mbox{Appendix~\ref{apx:Aint}}).}

We can see in figure~\ref{fig:VC_C0_tnd-OverAll-JFM}(\textit{a}) that the quantity of methylene blue dye in the drop initially decreases exponentially in time, as predicted by our Newton-cooling model (\ref{eq:NewtonCooling}) (plotted with a solid line).  {However, $\mathcal{M}_{MB}/\mathcal{M}_{MB,0}$}  decreases faster than our model prediction for $t/\tau_{\kappa}\gtrapprox 1$. The faster decrease at late times is due to noise in the experimental signal measured by the detection system. We  found that the detection system was not able to measure all the dye present in the drop, particularly at the edge where the drop is slightly thinner. This can be seen in figure~\ref{fig:Alldrops} where, at late time ($t\geq100$), the dye at the edge disappears more quickly than in the centre, even though the shape of the drop remains unchanged. To account for the dye missed by the detection system, we correct our experimental data as follows. 
 {We assume that in the part of the drop where the dye is still visible the spatially averaged concentration measured in this part is the same as throughout the entire drop. 
Then, assuming that this visible part  is the top of the drop, its shrinking volume in time is}
\begin{equation}
V_{vis}(t) = \frac{2}{3}\pi R^3 \(1 - \(1-\(\frac{R_{vis}(t)}{R}\)^2\)^{3/2}  \) - \pi \mbox{$R_{vis}$}^2(t) \(R^2-\frac{A_m}{\pi}\)^{1/2},
\end{equation}
 {where $R$ is the radius of curvature of the drop computed from its measured  volume $V_0$ and  maximum asymptotic area $A_m$, assuming a spherical cap shape. Therefore, the total corrected amount of dye in the drop is}
\begin{equation}\label{eq:corrdyedata}
m_{MB}(t) = \frac{\mathcal{M}_{MB}(t)}{V_{vis}(t)} V_0.
\end{equation}

In figure~\ref{fig:VC_C0_tnd-OverAll-JFM}(\textit{b}),  {we present the corrected normalised data $m_{MB}/m_{MB,0}$}. Our model for the missing dye seems to have correctly identified the source of discrepancy between the uncorrected data and the Newton-cooling model (plotted again with a solid line) observed in figure~\ref{fig:VC_C0_tnd-OverAll-JFM}(\textit{a}).
The ensemble-averaged data of all the different experiments, with film Reynolds number varying from $500$ to $1100$,  collapse with our model at all time $0\leq \hat{t} \leq 2.5$.  We find that the characteristic cleaning time ranges from $\tau_{\kappa}=37$ to $92\;$s and depends only on the ratio of the characteristic drop thickness $h_d=V_0/A_m$ to the convective mass transfer coefficient $\kappa$.

Therefore, the empirical model based on an analogy with Newton's law of cooling captures the overall physical processes controlling the mass transfer of a tracer out of a small droplet into a surrounding film flow at large film \Pns and for all drop \Pns. Moreover, the theoretical model based on fundamental physical principles developed in sections \ref{sec:transfilm} and \ref{sec:modeldropphase}, which also predicted an exponential decrease of the  spatially averaged concentration in the drop for $\hat{t}>0.1$ (see (\ref{eq:1ddiffNDsolavg}) and figure~\ref{fig:DropCmodel}\textit{b}), is also in agreement with this finding. 

It is interesting to note that the exponential decrease of the drop concentration seems independent of the drop \Pn, which ranges $0.04 \leq \Pe_d \leq 0.6$ (see figure~\ref{fig:t_d_difadv_Ped-OverAll}\textit{b}). This suggests that the range of validity of the theoretical model in sections \ref{sec:transfilm} and \ref{sec:modeldropphase} might extend beyond the drop diffusion dominated regime (regime D where $Pe_d\ll 1$). As we discussed at the end of section \ref{sec:modeldropphase}, the theoretical prediction (\ref{eq:1ddiffNDsolavg}) for $C_d$ might still be appropriate for $\Pe_d \approx 1$ (regime AD). 
Thus, for moderate drop \Pns $\Pe_d \approx 1$, we also have
\begin{equation}\label{eq:concregimeAD}
\hat{C}_d \approx \exp[-t/\tau_d],
\end{equation}
which is equivalent to the prediction of the Newton-cooling model (\ref{eq:Csolution}). In regime AD, since diffusive and advective processes are of the same importance, the characteristic transport time in the drop should be of the same order of both diffusive and advective time scales such that $\tau_d \approx \tau_{d,dif} \approx \tau_{d,adv} \approx \mbox{$h_d$}^2/D_d$. Advection processes in the drop could perhaps influence $\tau_d$ in the form of an enhanced effective diffusion coefficient larger than $D_d$.

\subsection{Mass transport in the drop}

\begin{figure}
	\centering
	\input{./t_d_difadv_etad-OverAlllog}
	\includegraphics[width=0.49\textwidth]{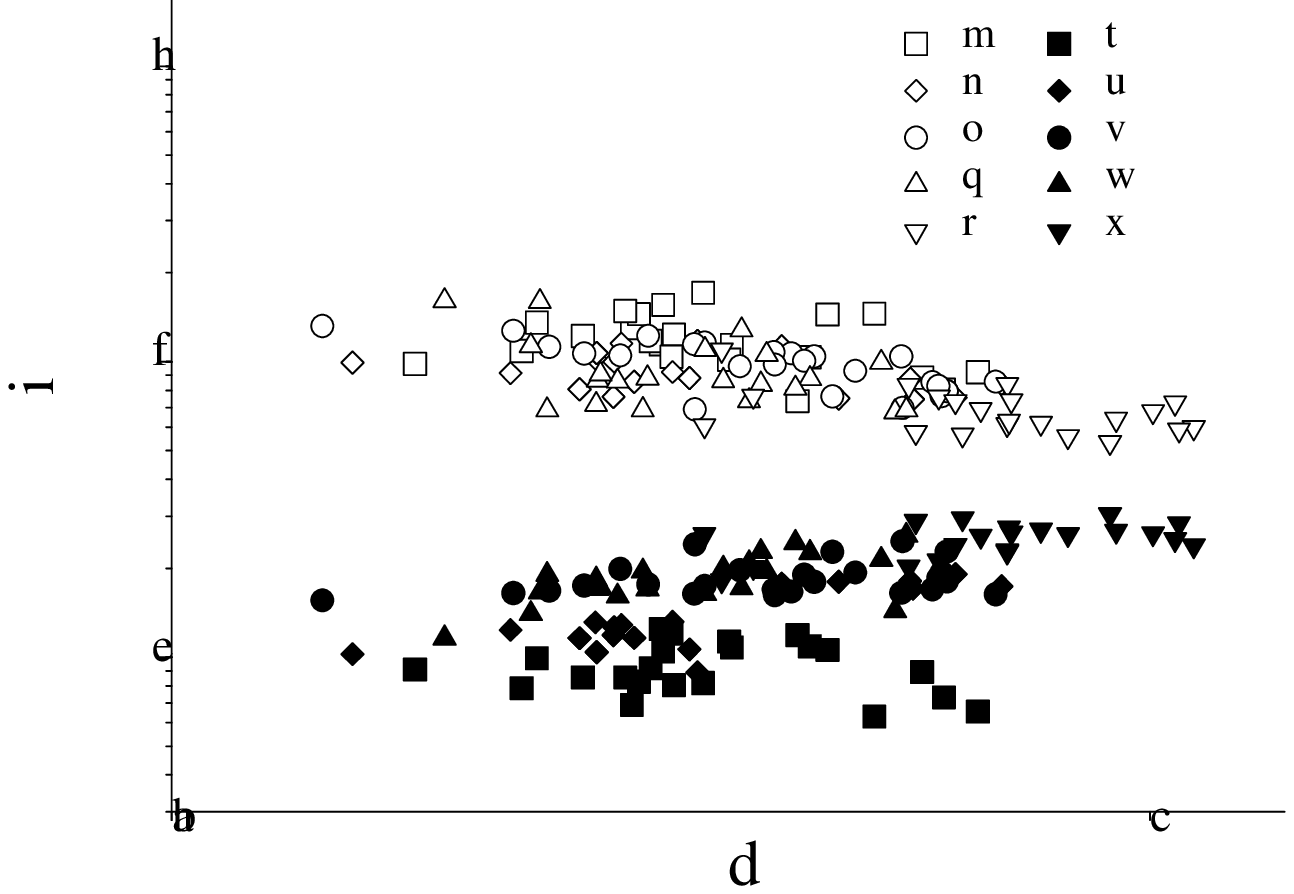}~
	\input{./Pe_d_etad-OverAll}
	\includegraphics[width=0.49\textwidth]{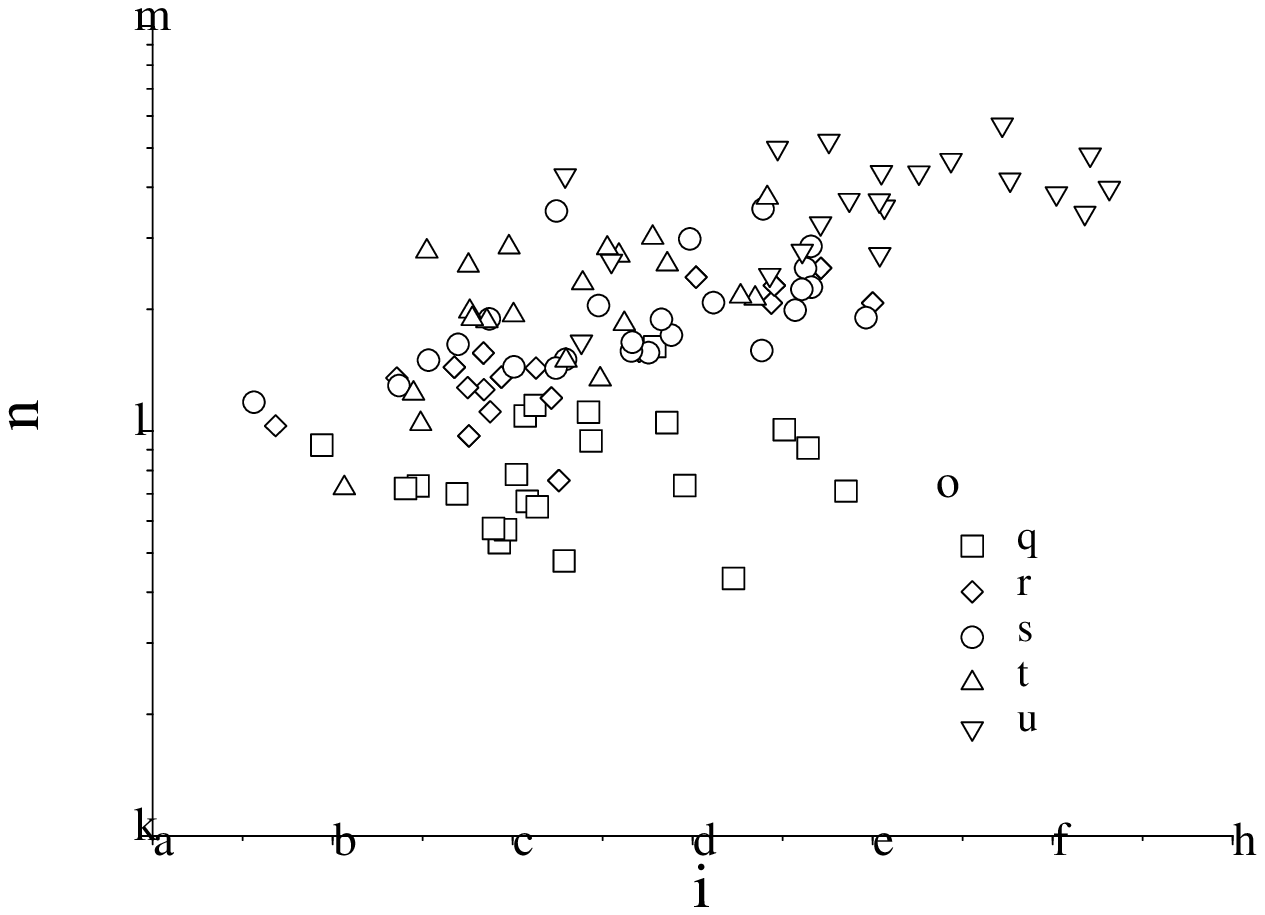}
	\caption{(\textit{a}) Normalised convective mass transfer time scales, $\tau_{\kappa}/\tau_{d,dif}$ and $\tau_{\kappa}/\tau_{d,adv}$ (plotted with open and closed symbols respectively), versus the drop aspect ratio $\eta_d$,  {on logarithmic axes}. (\textit{b}) Distribution of the drop \Pn $\Pe_d=\tau_{d,dif}/\tau_{d,adv}$ on a logarithmic axis versus the drop aspect ratio $\eta_d$. We plot the experimental data for all the $119$ drops of the $5$ sets of experiments (shown with different symbols).}
	\label{fig:t_d_difadv_Ped-OverAll}
\end{figure}

In figure~\ref{fig:t_d_difadv_Ped-OverAll}(\textit{a}), we plot  {on a vertical logarithmic axis} the ratio of the characteristic convective mass transfer time scale $\tau_{\kappa}=V_0/(\kappa A_m)$ with the drop characteristic diffusive time scale $\tau_{d,dif}=4 \mbox{$h_d$}^2/(\pi^2 D_d)$ (see (\ref{eq:tauddiff2})) using open symbols. The ratio of $\tau_{\kappa}$ with the drop characteristic advective time scale $\tau_{d,adv}=\mu_d L/(\mu_f \gamma h_d)$ (see (\ref{eq:taudadv})) is plotted using closed symbols. The horizontal logarithmic axis gives the drop aspect ratio $\eta_d=h_d/L$. 
In figure~\ref{fig:t_d_difadv_Ped-OverAll}(\textit{b}), we plot the drop \Pn $\Pe_d=\tau_{d,dif}/\tau_{d,adv}$ (see (\ref{eq:dropPn})) on a logarithmic axis against $\eta_d$.
In both figures we plot the experimental data of all the $119$ drops of the $5$ sets of experiments using different symbols.

As we can see in figure~\ref{fig:t_d_difadv_Ped-OverAll}(\textit{a}), $\tau_{\kappa}/\tau_{d,dif}$ is of the order of $1$, which means that for all the drops diffusion processes inside the drop have a large impact on the overall transport.  {This confirms our prediction \mbox{(\ref{eq:taukappaeqtauddif})}}. The scatter in the data, $0.5\leq \tau_{\kappa}/\tau_{d,dif}\leq 1.7$, is partially due to noise in the indirect estimation of $h_d$ in the computation of  $\tau_{d,dif}$. The drop height is estimated  from  $h_d=V/A$, where the volume of the drop is assumed constant and equal to $V_0$ and the area of the drop is chosen as the asymptotic area $A_m$ reached by the drop after deformation. 

The  ratio  $\tau_{\kappa}/\tau_{d,adv}$ is much smaller and ranges from $0.05$ to $0.3$. This shows that advection processes inside the drop have a negligible to small impact on the overall transport. We can also observe that $\tau_{\kappa}/\tau_{d,dif}$ tends to decrease with $\eta_d$ from approximately $1.5$ at $\eta_d \approx 0.1$ to $0.5$ at $\eta_d \approx 0.2$, while $\tau_{\kappa}/\tau_{d,adv}$ increases from approximately $0.1$ to $0.3$ in the same range.

The influence of the drop aspect ratio on the drop diffusion time is only through the drop height as $\tau_{d,dif} \propto \mbox{$h_d$}^2$, and hence $\tau_{d,dif} \propto \mbox{$\eta_d$}^2$. On the other hand, the drop advection time is inversely proportional to  the drop aspect ratio according to (\ref{eq:taudadv}), as thicker drops allow a larger internal drop velocity, whereas  {thinner} drops increase the distance of recirculation. If $\tau_{\kappa}$ was constant, we would have $\tau_{\kappa}/\tau_{d,dif} \propto 1/\mbox{$\eta_d$}^2$ and   $\tau_{\kappa}/\tau_{d,adv} \propto \eta_d$. The trends of both ratios in figure~\ref{fig:t_d_difadv_Ped-OverAll}(\textit{a}) seem to agree, although with  {some  scattering, with these predictions from our model}. However, $\tau_{\kappa}$ is not constant in our experiments and varies from $37$ to $92\;$s, which probably explains part of the scattering, in addition to the experimental error for $h_d$ discussed above.

All these results can be interpreted in the light of the drop \Pn, shown in figure~\ref{fig:t_d_difadv_Ped-OverAll}(\textit{b}), which ranges from approximately $0.04$ to $0.6$. At $\eta_d\approx 0.1$, where $\tau_{\kappa}/\tau_{d,dif}$ is of order $1$ and $\tau_{\kappa}/\tau_{d,adv}$ is much smaller, the drop \Pn is mostly in the diffusion dominated regime (regime D) where $Pe_d\ll 1$. On the other hand, as $\eta_d$ increases to $0.2$ there is a transition towards the advection--diffusion regime (regime AD) where $Pe_d$  increases to approach $1$. 

Although the range of our data set is not large enough to identify the asymptotic limit as $\eta_d \rightarrow 1$, it suggests that in this limit the drop \Pn increases and advection processes become more dominant for the transport within the drop. In this regime, the overall transport from the bulk of the drop to the bulk of the film flow would thus be mainly influenced by  drop advective processes, provided that we remain in the limit of negligible transport time in the film (\ie $\tau_f \ll \tau_d$).

\subsection{Mass transport in the film}\label{sec:mtFilm}

\begin{figure}
	\centering
	\input{./ShL2_ReL-OverAll-JFM}
	\includegraphics[width=0.49\textwidth]{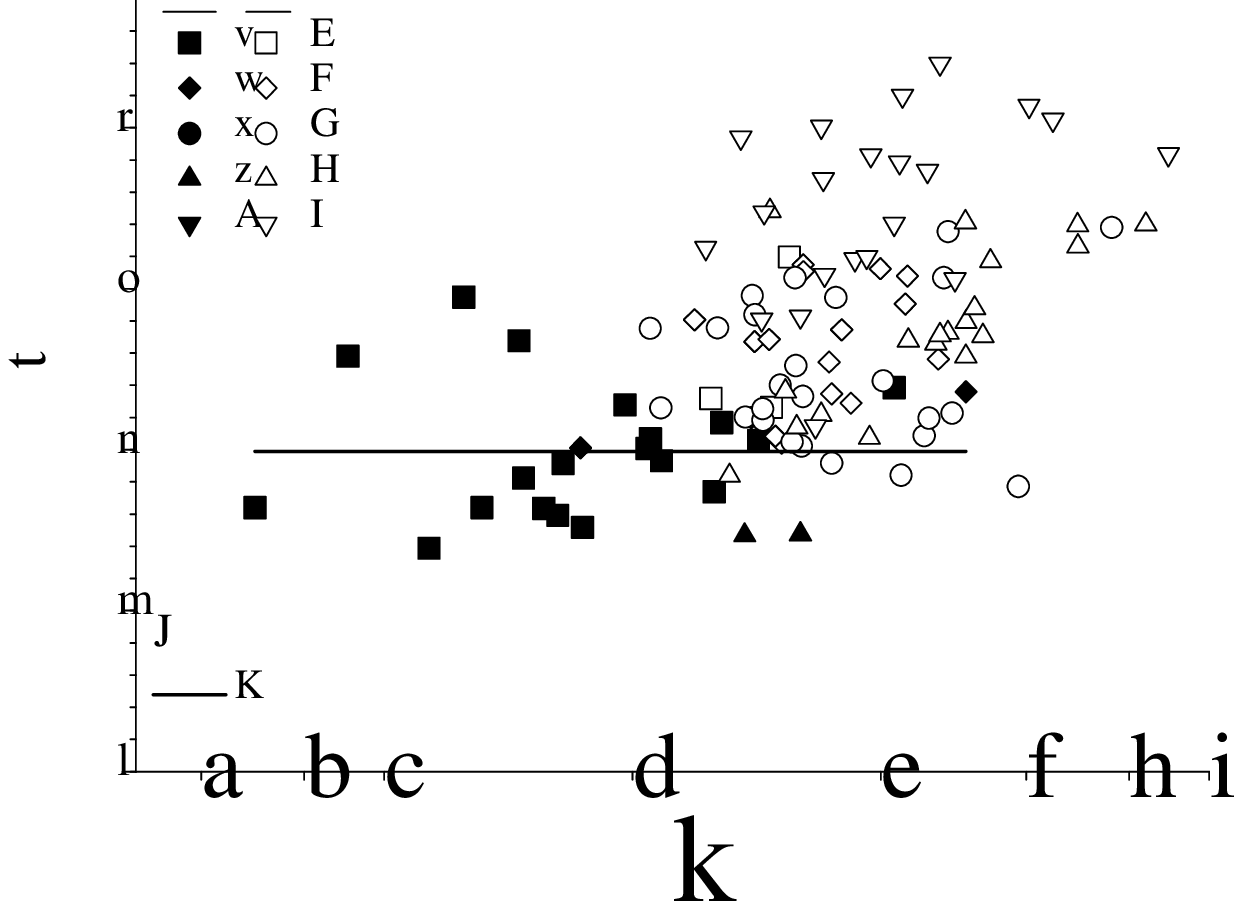}~
	\input{./ShL_Re_fbl-OverAll-JFM}
	\includegraphics[width=0.49\textwidth]{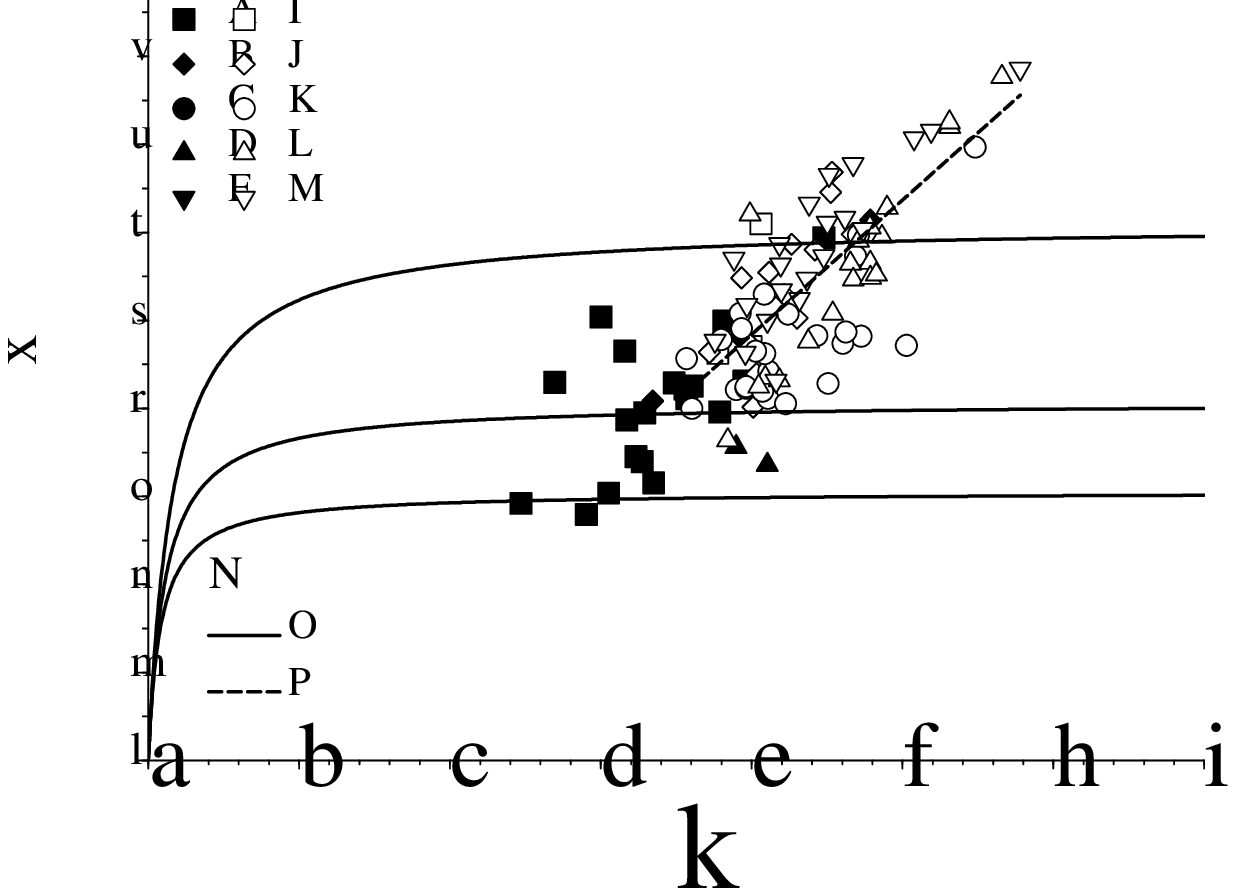} 
	\caption{(\textit{a}) Ratio of the experimentally measured spatially averaged Sherwood number $\overline{\Sh}=\kappa \ell/D$ with the theoretical prediction $\overline{\Sh}_{model,D}$  for regime D, $\Pe_d \ll 1$, (see (\ref{eq:ShvsRe13}))   against  {the  Reynolds} number $\Rey_L = \gamma L^2/\nu_f$ plotted on a logarithmic axis. The solid line is the average of the data for regime D.  {(\textit{b}) Distribution of $\overline{\Sh}$ versus $\Rey_{\delta}$. We plot the theoretical prediction \mbox{(\ref{eq:ShmodelD2})} for $\Pe_d \ll 1$ using three solid lines for $\eta_d=0.1$, $0.15$ and $0.2$ (from top to bottom). The dashed line corresponds to a least-squares linear fit of the data for $\Pe_d > 0.1$. In both graphs, closed symbols correspond to $\Pe_d\leq 0.1$ or the drop diffusion regime (D), and open symbols correspond to $\Pe_d > 0.1$ or the drop advection--diffusion regime (AD).}}
	\label{fig:ShL_ReL-OverAll-JFM}
\end{figure}

In figure~\ref{fig:ShL_ReL-OverAll-JFM}(\textit{a}), we show the ratio of the experimentally measured spatially averaged  Sherwood number $\overline{\Sh}=\kappa \ell/D$ (see (\ref{eq:defShkappa})) (with $\ell=\sqrt{A_m}$) to the theoretical prediction $\overline{\Sh}_{model,D}$ (see (\ref{eq:ShvsRe13}))  {for regime D. We plot data for all $119$ experiments, with closed symbols for  regime D ($Pe_d \leq 0.1$), and open symbols for  regime AD ($Pe_d > 0.1$). The ratio is plotted against $\Rey_L = \gamma L^2/\nu_f$ on a logarithmic axis, which varies from $7\times 10^3$ to $10^5$.
The} data for $\overline{Sh}/\overline{Sh}_{model,D}$ in regime D appears constant over the whole range $7\times 10^3\leq \Rey_L \leq 5\times 10^4$ and the average value, plotted with a solid line, is equal to $0.99$, with a standard deviation of $0.20$. This result validates our revised model (\ref{eq:ShvsRe13}) in the case of drops dominated by diffusion processes. In this model, the Sherwood number is not only dependent on  {$\Rey_L$ and $Sc_f$}, but also on the drop aspect ratio $\eta_d$, which varies from $0.1$ to $0.2$ approximately (see figure~\ref{fig:t_d_difadv_Ped-OverAll}), and the diffusivity ratio $\xi=D_f/D_d$ (constant in our experiments). We note that $\overline{Sh}_{model,D}$ does not have any fitting or empirical parameter, but is completely defined a priori, based on fundamental physical principles. 

In contrast, figure~\ref{fig:ShL_ReL-OverAll-JFM}(\textit{a}) shows that the Sherwood number for the drops in regime AD does not  {agree  well} with the theoretical prediction for regime D: $\overline{Sh}$ increases with $\Rey_L$ more rapidly than $\overline{Sh}_{model,D}$. Combining these results with the results for the drop \Pn shown in figure~\ref{fig:t_d_difadv_Ped-OverAll}(\textit{b}), we find that as  $Pe_d$ increases from approximately $0.1$ to $0.6$, the Sherwood number experiences a transition from regime D, described  {by}  (\ref{eq:ShvsRe13}), to another regime where $\Rey_L$ has more influence on $\overline{Sh}$.

 {We noted at the end of \mbox{section~\ref{sec:transfilm}} that the characteristic Reynolds number for the film Sherwood number $\overline{Sh}_f$ depends on the length of the drop as characteristic length scale $\Rey_L=\gamma L^2/\nu_f$ (see \mbox{(\ref{eq:SherNum1order})}). This result, emerging naturally from the advection--diffusion balance in the diffusive boundary layer, is consistent with past studies} \cite[][]{stone89,baines:james94,danberg08}.  {Nevertheless, the physical meaning of $\Rey_L$ is not entirely intuitive in the case of a film flow where the length of the drop $L$ can be greater than the film thickness $h$. In the  studies mentioned above, the mass transfer occurred in an infinite domain under a linear external shear flow. Thus, the length of the drop $L$ was the only natural characteristic length scale in the problem. In our problem, $\Rey_L$ appears to have an inconsistent  characteristic velocity as $U_L = \gamma L$ is larger than the film characteristic velocity $U_f=\gamma h/3$ in most experiments, and always much larger than the characteristic velocity in the  diffusive boundary layer $U_{\delta} = \gamma \delta$ since $\delta \ll L$.
This apparent physical inconsistency can be resolved if we substitute for $L$ in $\Rey_L$ using the estimation of the diffusive boundary layer thickness $\delta =(D_f L/\gamma)^{1/3}$ (see \mbox{(\ref{eq:deltafbl})}), thus we have}
\begin{equation}
\mbox{$\Rey_L$}^{1/3} \mbox{$\Sc_f$}^{1/3} = Pe_{\delta} = Re_{\delta} Sc_f,
\end{equation}
 {with the characteristic \mbox{\Pn} in the diffusive boundary layer}
\begin{equation}
Pe_{\delta} = \frac{\gamma \delta^2 }{D_f} \quad = \quad  \mbox{$Pe_L$}^{1/3} = \( \frac{\gamma L^2 }{D_f} \)^{1/3},
\end{equation}
 {and the associated Reynolds number}
\begin{equation}\label{eq:Refbl}
Re_{\delta} = \frac{\gamma \mbox{$\delta$}^2 }{\nu_f}.
\end{equation}
 {The prediction for the Sherwood number \mbox{(\ref{eq:SherNum1order})} then becomes a simple linear relationship}
\begin{equation}\label{eq:Shf_Pefbl}
\overline{Sh_f} = 0.766 Pe_{\delta} = 0.766 Re_{\delta} Sc_f.
\end{equation}
 {This relationship is mathematically equivalent to \mbox{(\ref{eq:SherNum1order})}, and should therefore apply to the well-mixed drop regime with $\overline{Sh_f}=\overline{Sh}$. It should also apply to the cases studied by} \cite{coutant:penski82}, \cite{baines:james94}, \cite{danberg08} and \cite{blount10}  {for the mass transfer from a drop constituted of a single species, and the case of heat transfer in a shear flow studied by} \cite{stone89}.  {The main difference with \mbox{(\ref{eq:SherNum1order})}  is that the physical interpretation of \mbox{(\ref{eq:Shf_Pefbl})} is clearer and more intuitive. Indeed, the characteristic length and velocity scales for the \mbox{\Pn} $Pe_{\delta}$ and the Reynolds number  $Re_{\delta}$ associated with the mass transfer are simply the diffusive boundary layer thickness $\delta$ and its characteristic velocity $U_{\delta}\approx \gamma \delta$. As \mbox{\cite{bejan13}} pointed out, for most flows, the local Reynolds number that is physically meaningful ``is based on the local longitudinal velocity scale (\mbox{\ie $U_{\delta}$}) and the local transverse dimension of the stream (\mbox{\ie $\delta$})''. The \mbox{\Pn} $Pe_{\delta}$ in \mbox{(\ref{eq:Shf_Pefbl})} represents the ratio of the longitudinal  ($x$) advective transport rate with the normal  ($y$) diffusive transport rate in the diffusive boundary layer. It naturally emerges from the dimensional analysis of the advection--diffusion equation (see \mbox{(\ref{eq:advdiff1})}) in the diffusive boundary layer, which states the balance of orders of magnitude}
\begin{equation}
 \frac{U_{\delta} \Delta C}{L} \sim D_f \frac{\Delta C}{\mbox{$\delta$}^2},
\end{equation}
 {(with $\Delta C$ a characteristic concentration variation) and thus}
\begin{equation}
1 \sim \frac{\delta}{L} \frac{U_{\delta} \delta}{D_f} = \eta_{\delta} Pe_{\delta},
\end{equation}
 {with $\eta_{\delta}=\delta/L \ll 1$ the slenderness ratio of the diffusive boundary layer. Therefore, $Pe_{\delta}$ characterizes the physics of the mass transfer at the small scale, \mbox{\ie} at the scale of the diffusive boundary layer $\delta$. It is mathematically equal to $\mbox{$Pe_L$}^{1/3}=(Re_L Sc_f)^{1/3}$, which is  a more accessible measure of the mass transfer as it is based on the main natural characteristic length  in the problem: the drop length $L$. This observation, although based on simple dimensional analysis, gives further justification for the scaling in the well-mixed model $\overline{Sh}\propto Pe_{\delta} = \mbox{$Pe_L$}^{1/3}$. Using \mbox{(\ref{eq:Shf_Pefbl})}, we can see that the prediction \mbox{(\ref{eq:ShvsRe13})} in regime D ($\Pe_d \ll 1$)  can also be modified to}
\begin{equation}\label{eq:ShmodelD2}
\overline{\Sh}_{model,D} \approx \frac{0.766 Pe_{\delta}}{1+ 0.350 \eta_d  \xi Pe_{\delta}}.
\end{equation}

 {In figure~\mbox{\ref{fig:ShL_ReL-OverAll-JFM}}(\textit{b}) we show the same experimental data for $\overline{Sh}$ as in figure~\mbox{\ref{fig:ShL_ReL-OverAll-JFM}}(\textit{a}), using linear axes, against $Re_{\delta}=\gamma \mbox{$\delta$}^2/\nu_f$. The theoretical prediction \mbox{(\ref{eq:ShmodelD2})} for $\overline{Sh}_{model,D}$ (plotted with three solid lines for $\eta_d=0.1$, $0.15$ and $0.2$, from top to bottom, and using the experimental values $\Sc_f=2000$ and $\xi=5/3$) is almost constant with $Re_{\delta}$ in the range $0.12\leq Re_{\delta} \leq 0.25$. In fact,  $\overline{Sh}_{model,D}$ is very close to its maximum asymptotic value, described in \mbox{(\ref{eq:Shmax})}: $\overline{\Sh}_{max}=\pi^{5/2}/(8\eta_d \xi)=12$, $8$ and $6$ for $\eta_d=0.1$, $0.15$ and $0.2$, respectively. The experimental data corresponding to the drop diffusive regime D  (plotted using closed symbols) appear rather constant and all within the three solid lines of $\overline{Sh}_{model,D}$, as expected from figure~\mbox{\ref{fig:ShL_ReL-OverAll-JFM}}(\textit{a}). On the other hand, the  data in regime AD,  plotted with open symbols, show a clear increasing trend with $\Rey_{\delta}$. We compute a least-squares linear fit of  the data in regime AD (plotted with a dashed line) using $\overline{\Sh} = K_1  \Sc_f \Rey_{\delta} + K_2$. We find the fitting parameters $K_1=0.032$ (standard deviation $0.003$) and $K_2=-2.5$ (standard deviation $1.3$). Although there is some scatter in our data (the Pearson correlation coefficient is $0.74$), our results suggest a linear relationship, as predicted by the well-mixed model  $\overline{\Sh}\propto \Rey_{\delta}$ in \mbox{(\ref{eq:Shf_Pefbl})} and its mathematical equivalent  $\overline{\Sh}\propto \mbox{$\Rey_{L}$}^{1/3}$ in \mbox{(\ref{eq:Shwellmixed})}.}

The apparent discrepancy between the coefficient of proportionality measured from the fit,  {$K_1=0.032$, and the coefficient of proportionality $0.766$ in equations  \mbox{(\ref{eq:Shwellmixed})} or \mbox{(\ref{eq:Shf_Pefbl})}} is due to a slightly different way of computing the Sherwood number. The experimental Sherwood number $\overline{Sh}=\kappa \ell/D_f$ uses the drop spatial average concentration $C_d$ as reference according to (\ref{eq:defShkappa}) and (\ref{eq:defkappa}). On the other hand, the well-mixed model in  \mbox{(\ref{eq:Shwellmixed})} or    {\mbox{(\ref{eq:Shf_Pefbl})}} uses the interfacial concentration $C_{f,i}$ as reference (see the definition of $\overline{Sh_f}$ in (\ref{eq:defSh})). Hence, there is a ratio $C_{f,i}/C_d$ between $\overline{Sh_f}$ and $\overline{Sh}$, as discussed in (\ref{eq:ShShf}). If the drops were  well-mixed in the experiments, then the concentration would be uniform throughout the drop with $C_{f,i} = C_d$ and the two Sherwood numbers $\overline{Sh_f}$ and $\overline{Sh}$ would be equivalent. However, the drops in  regime AD are unlikely to be completely uniform, as the characteristic transport time in the drop is still much longer than the transport time in the film: $\tau_d\approx 10$ to $100\;$s compared with  {$\tau_f\approx \tau_{\delta} \approx \SI{e-2}{}$ to $\SI{e-1}{s}$}. The drops in regime AD have  just started the transition from the non-uniform diffusion regime (D) to a well-mixed interior regime as advection processes become stronger. Nevertheless, they are still influenced by diffusion within the drop. This can be seen by the fact that the diffusive model (\ref{eq:ShvsRe13})  {or \mbox{(\ref{eq:ShmodelD2})}}, plotted with solid lines, predicts the correct order of magnitude for both regimes D and AD in figure~\ref{fig:ShL_ReL-OverAll-JFM}. In fact, the ratio  $C_{f,i}/C_d$ predicted by this model corresponds  very closely to the ratio between the fitting parameter $K_1$ and the coefficient of proportionality for the well-mixed model: 
\begin{equation}\label{eq:correctionPe}
\frac{K_1}{0.766} \approx 0.04 \approx \frac{C_{f,i}}{C_d} \approx <\frac{1}{1+ 0.350 \eta_d  \xi \Rey_{\delta} \Sc_f  }>.
\end{equation}
 {Here, $<\cdot >$ represents the ensemble average for the range of Reynolds numbers} $0.1\leq \Rey_{\delta}\leq 0.3$, and drop aspect ratios, $0.1 \leq \eta_d \leq 0.2$, studied.

We can also observe in  figure~\ref{fig:ShL_ReL-OverAll-JFM}(\textit{b}) that most of the data in regime D (closed symbols) correspond to lower  Reynolds numbers with  {$\Rey_{\delta}\leq 0.2$}. In contrast, all the data in regime AD (open symbols) are found for  {$\Rey_{\delta}\geq 0.2$}. This is due to the coupling of the velocity field in the film with the velocity field in the drop. The recirculation velocity in the drop is directly proportional to the film characteristic shear rate $U_d\propto \gamma$ according to (\ref{eq:Ud}). Since  {$\Rey_{\delta}$} is also based on $\gamma$, the drop \Pn is  indirectly related to the  film flow through the continuity of the stress at the interface.

\begin{figure}
	\centering
	\input{./Sh_Pe_f_bl-Others-JFM}
	\includegraphics[width=0.9\textwidth]{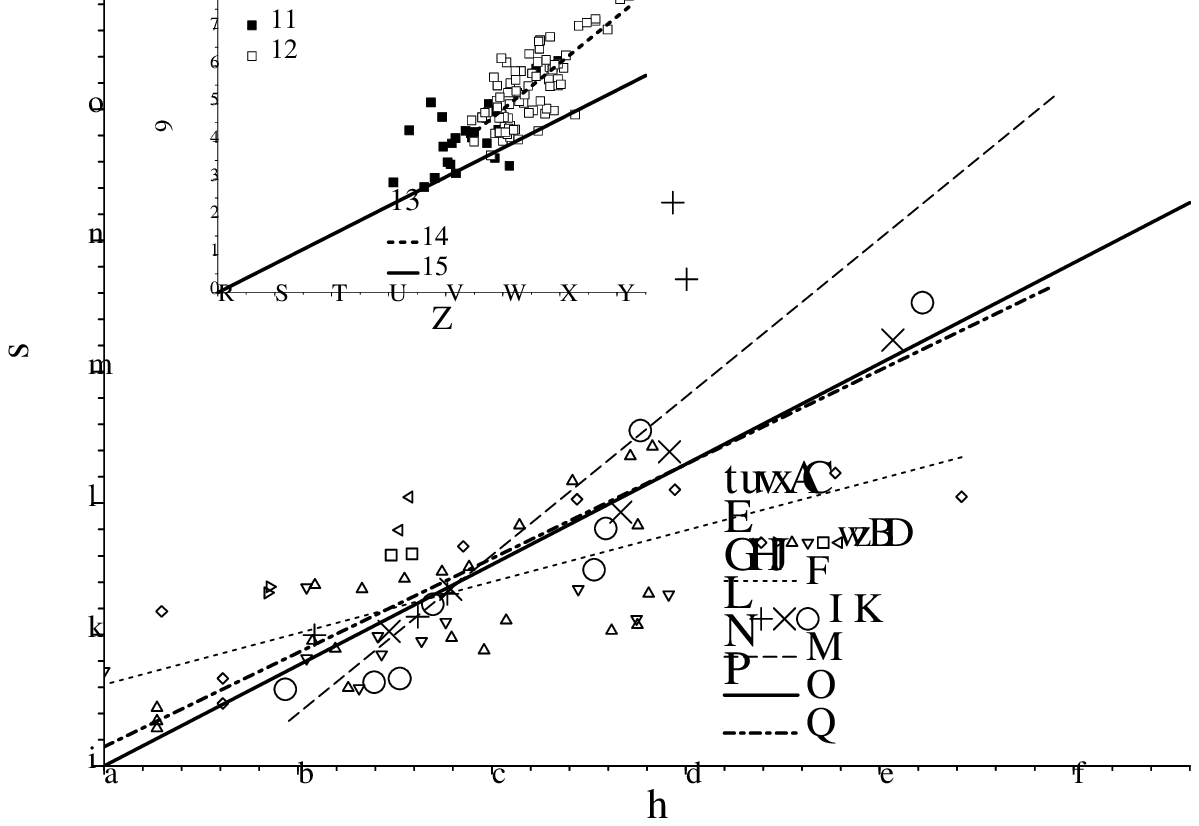}~
	\caption{ {Distribution of experimental data $\overline{\Sh}$ versus \mbox{\Pn} $Pe_{\delta}$. The linear theoretical prediction for the well-mixed regime \mbox{(\ref{eq:Shf_Pefbl})} is plotted with a thick solid line in both graphs. (\textit{a}) Data of \mbox{\cite{coutant:penski82}} for the evaporation in air of drops of: $\diamond$ dimethylformamide on glass;  $\triangleright$ water, {\scriptsize $\triangle$} dimethylformamide, $\triangledown$ ethylbenzene, {\tiny $\square$} mesitylene, and $\triangleleft$ butyl alcohol, all on Teflon. The dotted line correspond to a least-squares linear fit of the data of \mbox{\cite{coutant:penski82}}. Data of \mbox{\cite{danberg08}} for the evaporation in air of  drops of agent HD (chemical formula C$_4$H$_8$Cl$_2$S) on glass substrate at different temperatures: $+$ $15^{\circ}$C, $\times$ $35^{\circ}$C, and {\scriptsize $\bigcirc$} $50^{\circ}$C. The dashed line correspond to a least-squares linear fit of the data of \mbox{\cite{danberg08}}. The thick dot-dash line correspond to all the data in (\textit{a}). (\textit{b}) Same data as in figure~\mbox{\ref{fig:ShL_ReL-OverAll-JFM}} for $\overline{\Sh}$, plotted with closed squares for regime D and open squares for regime AD. The mean correction \mbox{(\ref{eq:correctionPe})} has been applied to compute $Pe_{\delta}$.  The dashed line correspond to a least-squares linear fit in regime AD: $\overline{\Sh}=1.2 Pe_{\delta} -2.5$.}}
	\label{fig:Sh_Pe_f_bl-Others-JFM}
\end{figure}

 {In figure~\mbox{\ref{fig:Sh_Pe_f_bl-Others-JFM}} we show the evolution of the experimental data of \mbox{\cite{coutant:penski82}} and \mbox{\cite{danberg08}} (\textit{a}) and our data (\textit{b}) for the Sherwood number $\overline{Sh}$ with respect to the \mbox{\Pn} $Pe_{\delta}$. Plotted using linear axes, we compare these data with the linear alternative model we proposed in \mbox{(\ref{eq:Shf_Pefbl})}, $\overline{\Sh}=0.766 Pe_{\delta}$, in the well-mixed or saturated regime (plotted with a thick solid line in both graphs).}

 {The data of \mbox{\cite{coutant:penski82}} \mbox{\cite[also presented in][]{baines:james94}} in figure~\mbox{\ref{fig:Sh_Pe_f_bl-Others-JFM}}(\textit{a}) correspond to the evaporation of homogeneous drops of various substances: water, dimethylformamide, ethylbenzene, mesitylene and butyl alcohol on glass or Teflon substrates (see caption for corresponding symbols). A least-squares linear fit of their data (plotted with a dotted line) assuming $\overline{\Sh}=K_1 Pe_{\delta}+K_2$ gives: $K_1=0.39$ (standard deviation $0.06$) and $K_2=3.1$ (standard deviation $0.6$). The Pearson correlation coefficient is $0.67$ thus suggesting a weak linear correlation. 
The data of \mbox{\cite{danberg08}} correspond to the evaporation of homogeneous drops of agent HD (chemical formula C$_4$H$_8$Cl$_2$S) on glass substrate at different temperatures: $15^{\circ}$C, $35^{\circ}$C, and $50^{\circ}$C (see caption for corresponding symbols). A least-squares linear fit of the data  (plotted with a dashed line) gives: $K_1=1.21$ (standard deviation $0.14$) and $K_2=-4.0$ (standard deviation $1.7$). The  correlation coefficient is $0.90$, thus suggesting a strong linear correlation.}

 {We also plot using a dot--dash line in figure~\mbox{\ref{fig:Sh_Pe_f_bl-Others-JFM}}(\textit{a}) a least-squares linear fit of the combined data sets of \mbox{\cite{coutant:penski82}} and \mbox{\cite{danberg08}}. We find $K_1=0.72$ (standard deviation $0.14$), which is very close to the theoretical prediction of $0.766$, and $K_2=0.7$ (standard deviation $0.7$), which is also close to the theoretical prediction of $0$. The  correlation coefficient is $0.76$. Although we recognise the correlation coefficient is not very high, this finding supports the linear alternative model for the well mixed regime $\overline{\Sh}=0.766 Pe_{\delta}$. Our data, corrected using \mbox{(\ref{eq:correctionPe})}, are also shown in figure~\mbox{\ref{fig:Sh_Pe_f_bl-Others-JFM}}(\textit{b}). The dashed line corresponds to a least-squares linear fit in regime AD (data plotted with open squares): with $K_1=1.2$ (standard deviation $0.12$) and $K_2=-2.5$ (standard deviation $1.3$). The  correlation coefficient is $0.74$. Our data are   in broad agreement with the model due to the difficulty to correct them using \mbox{(\ref{eq:correctionPe})} related to the problem of reference concentration between $C_{f,i}$ and $C_d$ discussed above.}

 {The scattering in all the datasets presented in figure~\mbox{\ref{fig:Sh_Pe_f_bl-Others-JFM}}  shows the difficulty of convective mass transfer measurements. This difficulty is even greater in the case we studied due to the impact of the mass transport  in the interior of the non-homogeneous drop. Nevertheless, the analysis of our experimental data and past experimental data shows two main findings. Firstly, the mass transfer from a well-mixed or saturated drop follows the physically intuitive linear relationship $\overline{\Sh}\propto Pe_{\delta}= \gamma \mbox{$\delta$}^2/D_f$, which is mathematically equivalent to the well-known result $\overline{\Sh}\propto \mbox{$Pe_L$}^{1/3}=(\gamma L^2/D_f)^{1/3}$. Secondly, in the case of a non-homogeneous drop the Sherwood number depends on the drop \mbox{\Pn}. At low drop \mbox{\Pns} we find $\overline{\Sh}\propto Pe_{\delta}/(1+0.350 \eta_d \xi Pe_{\delta})$. As the drop  \mbox{\Pn} increases towards unity, a transition towards the well-mixed or saturated model is suggested.
}

\section{Conclusion}\label{sec:conc}

We have studied the convective mass transfer from a small isolated viscous droplet into a thin gravity-driven film. We analysed transport in the case of very large film \Pns, $\Pe_f \approx 10^6$, small to moderate drop \Pns,  {$\Pe_d\approx 10^{-2}$} to $1$, and in the limit of very small characteristic transport times in the film compared with characteristic transport times in the drop, $\tau_f \ll \tau_d$. The results of the present study have important implications for cleaning and decontamination applications. They show a quantitative validation of existing numerical and theoretical models in the case of a well-mixed or saturated drop. Moreover, they provide a new theoretical model in the case where the material to be removed from the droplet is in dilute concentration.

We found that an empirical model based on Newton's law of cooling can predict very accurately the exponential decrease in time of the tracer concentration in the drop. This model is valid for all the drop and film \Pns studied. We believe this model is particularly useful to many  applications owing to its simplicity, as it relies on very few parameters. The overall transport characteristic time of the model can be computed knowing only the constant drop volume, the asymptotic drop surface area and a convective mass transfer coefficient, such that $\tau_{\kappa}=V_0/\(\kappa A_m\)$.

To compute the convective mass transfer coefficient, or its non-dimensional equivalent the Sherwood number, using the properties of the flow, a more sophisticated model based on the advection--diffusion equation is needed. In the case of a well-mixed or saturated drop, we have found the same result as previous studies: $\overline{\Sh} = 0.766 \mbox{$\Rey_L$}^{1/3}\mbox{$\Sc_f$}^{1/3}$, with the  Reynolds number $\Rey_L=\gamma L^2/\nu_f$ and the film Schmidt number $ \Sc_f=\nu_f/D_f$. This result relies only on properties external to the drop.  {We also show that this result is mathematically equivalent to a simpler  relationship $\overline{\Sh} = 0.766 \Rey_{\delta} \Sc_f$, with $\Rey_{\delta}=\gamma \mbox{$\delta$}^2/\nu_f$. The first relationship uses the natural and directly accessible length scale in the problem: the drop length $L$. The second relationship is based on the local physical characteristic  length and velocity scales of the diffusive boundary layer,  $\delta$ and $\gamma \delta$ respectively. It states that the mass transfer coefficient is linear with the ratio of the diffusive rate across the diffusive boundary layer to the advective rate along the  drop surface: $\overline{\Sh} \propto Pe_{\delta}=Re_{\delta} Sc_f$.}

On the other hand, in the case of a dilute concentration of tracer in the drop, which is typical of decontamination applications where several species can be mixed in the drop, we showed the strong influence of transport within the drop. In the limit  $\tau_f \ll \tau_d$, we identified two regimes for the overall mass transfer: regime D with $Pe_d \ll 1$, dominated by diffusion inside the drop; and regime AD with $Pe_d \approx 1$, where both advection and diffusion processes inside the drop are important.

In regime D, we have proposed a theoretical model, based on fundamental principles, for the transport from the bulk of the drop into the bulk of the film. In the drop, we solve a one-dimensional time-dependent diffusion equation. We couple this equation with a quasi-steady advection--diffusion equation in the film diffusive boundary layer. The crucial point pertains to the fact that the diffusive boundary layer can be considered quasi-steady owing to $\tau_d \gg \tau_f$. Effectively, the slow time dependence of the mass transfer in the drop phase impacts the film phase only through the time-dependent interfacial concentration, which couples transport between the two phases.

This theoretical model supports the Newton-cooling model because it predicts the exponential decrease observed experimentally. We measured that the overall time scale of the mass transfer $\tau_{\kappa}$ is approximately equal to the drop diffusion time scale predicted by the theoretical model $\tau_{d,dif}=\mbox{$h_d$}^2/D_d$. This confirms that in regime D the overall mass transfer is  limited by  transport within the drop. 
Remarkably, the Newton-cooling model and the theoretical model also agree well with experimental data at $Pe_d \approx 1$ (regime AD). Both the exponential decrease in time and the characteristic transport time $\tau_{\kappa} = \tau_{d,dif}$ match between the models and the experiments. This shows that there is probably a  smooth transition in the dynamics of the mass transfer from regime D to AD as the drop \Pn increases. 

In regime D, the theoretical  model  predicts a correction for the Sherwood number compared with the well-mixed or saturated case  {depending on $Pe_{\delta}$}:
\begin{equation}\label{eq:Shconc}
\overline{Sh} \approx \frac{0.766 Pe_{\delta}}{1+ 0.350 \eta_d  \xi Pe_{\delta}},
\end{equation}
which now takes into account the drop aspect ratio $\eta_d = h_d/L$ and the diffusivity ratio $\xi=D_f/D_d$. The prediction for the well-mixed or saturated case can be recovered in the limit  {$0.350 \eta_d  \xi Pe_{\delta} \ll 1$}. Therefore, our new model can apply to both saturated drops (\ie with a single species) or non-uniform drops with a time-dependent concentration. We can also note that in the other limit,  {$0.350 \eta_d  \xi Pe_{\delta} \gg 1$},  we obtain a maximum asymptotic value for the Sherwood number, $\overline{Sh}_{max} = \pi^{5/2}/(8 \eta_d\xi)$, which is independent of the \Pn.

The experimental results agree remarkably well with the theoretical prediction for the Sherwood number (\ref{eq:Shconc}) at $Pe_d \leq 0.1$. We found that at the range of Reynolds numbers studied  {$Re_{\delta} \geq 0.1$}, the Sherwood number is close to its asymptotic limit $\overline{Sh}_{max}$. As the drop \Pn increases to $Pe_d \approx 1$, our theoretical model still predicts the correct order of magnitude for $\overline{Sh}$, although the influence of the Reynolds number increases.  {Our data suggest a linear correlation between $\overline{Sh}$ and $Pe_{\delta}$, as predicted by the well-mixed model.} However, we noted that there was still a strong gradient within the drop between the bulk concentration and the interfacial concentration. This also suggests that as $Pe_d$ increases, there is a smooth transition for the Sherwood number from $\overline{Sh} \approx \overline{Sh}_{max}$ in regime D to  {$\overline{Sh} \propto Pe_{\delta}$} in regime AD.

We believe that our model has captured the main physical processes of the convective mass transfer and the assumptions made were physically meaningful in the regimes considered. Our model is not only applicable for the well-mixed regime, already studied in previous papers, but also for the non well-mixed regime with transport within the drop limiting the overall mass transfer. In this regime, the model gives excellent results at low drop \Pns and reasonable estimates as $Pe_d$ approaches $1$. However, our theoretical model does not account for the regime of large drop \Pns $Pe_d \gg 1$. As the drop \Pn increases and advection processes become important inside the drop, the  recirculation flow might have a strong impact on the mass transfer. It is not exactly clear how the Sherwood number evolves when $Pe_d\gg 1$. In this advection dominated regime, transport should be described by a time-dependent advection--diffusion equation in the drop, coupled with a quasi-steady advection--diffusion equation in the film. In this regime, transport in the drop might also be fully three-dimensional. This can be a challenge for future studies.

\paragraph{Acknowledgments}
We wish to thank D. Page-Croft and the technicians of the GK Batchelor Laboratory at the Department of Applied Mathematics and Theoretical Physics, Cambridge. We are grateful to F. Bartholomew, F. Yuen and M. Etzold from the BP Institute, Cambridge, for their help in conducting contact angle and viscosity measurements. J. R. L. wishes to thank his colleagues P. Luzzatto-Fegiz and F. Peaudecerf for fruitful discussions. 
J. R. L. acknowledges financial support from Magdalene College, Cambridge, through a Nevile Research Fellowship in Applied Mathematics. This material is based upon work supported by the Defense Threat Reduction Agency under Contract No. HDTRA1-12-D-0003-0001.

\appendix

\section{Rheological study of the drop phase}\label{apx:RheoDrop}

  {We conducted a rheological study of the $2$\%~wt Natrosol  hydroxyethylcellulose (HEC) 250HHR polymer solution. The data were obtained with a Bholhin CVO rheometer (Malvern) using a cone and plate system ($40\:$mm in diameter and $2^{\circ}$ angle). We took measurements for the range of shear rate $\gamma$ from $0.1$ to $1000\:$s$^{-1}$, starting from the lowest shear rate, and with a $30\:$s plateau at each data point. We can fit our data with the Carreau-Yasuda model \mbox{\cite[see \eg][]{gijsen99}}, which relates the dynamic viscosity of a shear-thinning fluid to the shear rate such that}
\begin{equation}\label{eq:carreauyasuda}
\mu = \mu_{\infty} + \(\mu_0-\mu_{\infty}\)\( 1 + \( \lambda \gamma\)^{\alpha} \)^{\(n-1\)/\alpha},
\end{equation}
  {where  $\mu_{\infty}$ is the viscosity as  $\gamma \to \infty$,  $\mu_0$ the viscosity as  $\gamma \to 0$, $\lambda$  the relaxation time, $\alpha$  the Yasuda coefficient and  $n$ the power index. We can obtain these parameters for the $2$\%~wt HEC 250HHR polymer solution using a  least-squares fit of the experimental data (the typical standard deviation between the data and the fit is less than $0.004\;$Pa$\;$s): $\mu_{\infty}=0\;$Pa$\;$s, $\mu_0=212.3\;$Pa$\;$s, $\lambda=0.90\;$s, $\alpha=0.40$ and $n=6.49 \times 10^{-2}$. Our rheological data are in agreement with the manufacturer's measurements.}

\section{Mass transfer with a time-dependent interface area}\label{apx:Aint}

From our experimental observations, we assume an empirical model for the time evolution of the drop--film interface area, such that
\begin{equation}\label{eq:Aint}
A = \( A_m-A_0 \) \(1 - e^{-t/\tau_{d,def}}\) +A_0,
\end{equation}
  {with $A_0$ the initial interface area}. Replacing \mbox{(\ref{eq:Aint})} into \mbox{(\ref{eq:NewtonCooling})}, we find
\begin{equation}\label{eq:Csolution2}
\hat{C}_d = \exp{\[-\varphi\( \hat{t} + \psi\(1-\frac{1}{\varphi}\) \(e^{-\hat{t}/\psi} -1\) \)\]},
\end{equation}
  {where we use the non-dimensional concentration and time described in \mbox{(\ref{eq:normCt})}. Based on \mbox{(\ref{eq:kappadef})}, we define the characteristic time scale of the overall mass transfer using the initial drop area}
\begin{equation}
\tau_{\kappa} = \frac{V_0}{\kappa A_0}.
\end{equation}
  {We have introduced in \mbox{(\ref{eq:Csolution2})} two important parameters: the drop deformation factor and the ratio between the deformation time and the overall mass transfer time scale, respectively,}
\refstepcounter{equation}
$$
\varphi = \frac{A_m}{A_0}, \quad \psi = \frac{\tau_{d,def}}{\tau_{\kappa}},
\eqno{(\theequation{\textrm{\textit{a,b}}})}
$$
  {As expected intuitively, \mbox{(\ref{eq:Csolution2})} implies that both an increase in the drop area, $\varphi>1$, and an increase in the rate of change of the area compared with the mass transfer time scale, $\psi<1$, tend to decrease the drop concentration $C_d$ more rapidly.}

\section{Diffusivity of methylene blue dye in the drop and film phases}\label{apx:DiffExp}

  {We  conducted  experiments to measure the diffusivity of methylene blue in the two phases: water and the $2$\%~wt HEC polymer solution, as well as across the interface between the two phases. 
To suppress any advection processes and measure the diffusion in a one-dimensional domain, we conducted the diffusion experiments in capillary tubes (borosilicate glass) of length $150\:$mm, inner diameter $1.5\:$mm and outer diameter $3\:$mm. We filled one part of each tube with either tap water or the polymer mixture and some methylene blue. The other part of the tube contained either water or the polymer solution without any dye. 
The diffusion of the dye was monitored by measuring the change of concentration in time and space using the dye attenuation technique explained in \mbox{section~\ref{sec:xpproc}}.}

  {We conducted simultaneously four experiments in a room at $20^{\circ}$C, and with the solutions  at $\textrm{pH}=7$. The capillary tubes were oriented vertically. In \mbox{table~\ref{tab:diffusion}} we describe the composition of the top and bottom phases in each tube, at the onset of the experiment. The concentration of methylene blue (MB) was set at $C_0 = \SI{0.1}{kg. m^{-3}}$ ($0.01$\%~wt), so that the concentration in the initially undyed phase remained in the linear regime of the camera response throughout the experiment.}

\begin{table}
	\begin{minipage}{\textwidth}
		\centering
		\begin{tabular}{@{}lcccc@{}}
			\hline \\
			{Tube} & {T1} & {T2} & {T3} & {T4}\\[3pt]
			Top phase & water & water & polymer & polymer \\
			Bottom phase & water + MB & polymer + MB & water + MB & polymer + MB  \\
			$D \times 10^{10}$ (m$^2\:$s$^{-1}$) & $5$ & $4$ & $3$ & $3$ \\
			$C_0 \times 10^{4}$ (\%~wt) & $0.9$ & $1.2$ & $2.5$ & $1.1$ \\ \\ \hline
		\end{tabular}
	\end{minipage}
	\caption{Composition of the top and bottom phases in each capillary tube, at the onset of the diffusion experiment, and the measured diffusion coefficients.}
	\label{tab:diffusion}
\end{table}

  {To obtain the diffusion coefficients from the experimental measurements with the dye attenuation technique, we can assume that the concentration of methylene blue $C_{MB}$ follows the one-dimensional diffusion equation in both phases $\p C_{MB}/\p t = D \p^2 C_{MB}/\p x^2$, where $D$  is the constant diffusion coefficient of methylene blue, $t$ time (set at $0$ at the onset of the experiment) and $x$ the along-tube distance from the interface (increasing positively in the phase initially undyed). The length of the tube was very large compared with the typical diffusion distance. We assume that the diffusion of the polymer is negligible (our experimental observations confirm this assumption). The solution of the one-dimensional diffusion equation for a step function with initial conditions $ C_{MB}(x,t=0)=C_0$ for all $x<0$  and $ C_{MB}(x,t=0)=0$  for all  $x>0$ is}
\begin{equation}\label{eq:diffmodel}
\frac{C_{MB}}{C_0} = \frac{1}{2} \textrm{erfc}\[\frac{x-x_0}{\sqrt{4 D \(t-t_0\)}} \],
\end{equation}
  {with $\textrm{erfc}$  the complementary error function, and $x_0$  and  $t_0$ the spatial and temporal virtual origins, respectively.
To compute the diffusion coefficients $D$ and the unknown parameters  $x_0$, $t_0$ and $C_0$, we fit the one-dimensional diffusion model with the experimental data using a  least-squares fit. The diffusion coefficients of each experiment and the  fitting values for $C_0$ are presented in \mbox{table~\ref{tab:diffusion}}. 

We can see that the values of the diffusivity depends on the phase composition. The diffusivity in water is $D=5 \times 10^{-10}\:$m$^2\:$s$^{-1}$, and $D=3 \times 10^{-10}\:$m$^2\:$s$^{-1}$ for the polymer solution. Our measurements are consistent with the measurements of \mbox{\cite{sedlacek13}} who found $D=8.4 \times 10^{-10}\:$m$^2\:$s$^{-1}$ in pure water at $25^{\circ}$C and $D=2.9 \times 10^{-10}\:$m$^2\:$s$^{-1}$ in hydrogel solutions  at $25^{\circ}$C. The smaller diffusivity in the drop phase (constituted mainly of water) could be due to electro-chemical interactions between the methylene blue dye ions and some parts of the HEC polymer chain. The data are more difficult to interpret when the two phases are different. The transfer condition of the methylene blue ions at the interface might also be affected by electro-chemical interactions with the polymer.}

\bibliography{bibliography}
\bibliographystyle{jfm}

\end{document}